%% file: paper.tex
\documentclass[useAMS,usenatbib,a4paper]{mn2e}

\usepackage{aas_macros}
\usepackage{graphicx}
\usepackage{amsmath}
\usepackage{xspace}
\usepackage{lineno}
\usepackage{ifthen}
\usepackage{lscape}
\usepackage{placeins}
\newcommand{\ha}{\mbox{H$\alpha$}\xspace}
\newcommand{\kms}{\mbox{km\,s$^{-1}$}\xspace}
\newcommand{\teff}{$T_\text{eff}$\xspace}    
\newcommand{\logg}{$\log g$\xspace}
\newcommand{\feh}{[Fe/H]\xspace}

\newcommand{\chisq}{$\chi ^2$\xspace}

\title[An M-dwarf binary from HATSouth]{A $0.24+0.18 \,M_\odot$  double-lined eclipsing binary from the HATSouth survey
\thanks{The HATSouth network is operated by a
collaboration consisting of Princeton University (PU), the Max Planck
Institut f\"ur Astronomie (MPIA), the Australian National University
(ANU), and the Pontificia Universidad Cat\'olica de Chile (PUC). The
station at Las Campanas Observatory (LCO) of the Carnegie Institution
is operated by PU in conjunction with PUC, the station at the High
Energy Spectroscopic Survey (HESS) site is operated in conjunction
with MPIA, and the station at Siding Spring Observatory (SSO) is
operated jointly with ANU. This paper includes data gathered with the 6.5 meter Magellan Telescopes located at Las Campanas Observatory, Chile.
}}

\author[G.~Zhou et al.]
{\parbox{\textwidth}
{G.~Zhou$^{1}$\thanks{E-mail: \texttt{george.zhou@anu.edu.au}},
 D.~Bayliss$^{1}$,  
J.~D.~Hartman$^{2}$, 
M.~Rabus$^{3,4}$, 
G.~\'A.~Bakos$^{2}$\thanks{Alfred P.~Sloan Research Fellow}\thanks{Packard Fellow}, 
A.~Jord\'an$^{3}$,  
R.~Brahm$^{3}$, 
K.~Penev$^{2}$, 
Z.~Csubry$^{2}$, 
L.~Mancini$^{4}$, 
N.~Espinoza$^{3}$, 
M.~de~Val-Borro$^{2}$,
W.~Bhatti$^{2}$,
S.~Ciceri$^{4}$, 
T.~Henning$^{4}$, 
B.~Schmidt$^{1}$,
S.~J.~Murphy$^{5}$,
R.~P.~Butler$^{6}$,
P.~Arriagada$^{6}$,
S.~Shectman$^{7}$,
J.~Crane$^{7}$,
I.~Thompson$^{7}$,
V.~Suc$^{3}$ 
and R.~W.~Noyes$^{8}$\vspace{0.4cm}}\\
\parbox{\textwidth}{
$^{1}${Research School of Astronomy and Astrophysics, Australian National University, Canberra, ACT 2611, Australia}\\
$^{2}${Department of Astrophysical Sciences,
	Princeton University, NJ 08544, USA}\\
$^{3}${Instituto de Astrof\'isica, Pontificia Universidad
 Cat\'olica de Chile, Av.\ Vicu\~na Mackenna
   4860, 7820436 Macul, Santiago, Chile}\\
$^{4}${Max Planck Institute for Astronomy, K\"{o}nigstuhl 17,
69117 -- Heidelberg, Germany}\\
$^{5}${Astronomisches Rechen-Institut, Zentrum f\"{u}r Astronomie der Universit\"{a}t Heidelberg, D-69120 Heidelberg, Germany}\\
$^{6}${Department of Terrestrial Magnetism, Carnegie Institution
of Washington, 5241 Broad Branch Road NW, Washington, DC
20015-1305, USA}\\
$^{7}${The Observatories of the Carnegie Institution of Washington, 813 Santa Barbara Street, Pasadena, CA 91101, USA}\\
$^{8}${Harvard-Smithsonian Center for Astrophysics, 60 Garden St, Cambridge, MA, USA}}}

\begin{document}
%\linenumbers

\date{Accepted, 8 May 2015. Received, 8 May 2015; in original form, 28 November 2014}

\pagerange{\pageref{firstpage}--\pageref{lastpage}} \pubyear{2015}

\maketitle

\label{firstpage}

\begin{abstract}

We report the discovery and characterisation of a new M-dwarf binary, with component masses and radii of $M_1 = 0.244_{-0.003}^{+0.003} \, M_\odot$, $R_1 = 0.261_{-0.009}^{+0.006} \, R_\odot$,  $M_2 = 0.179_{-0.001}^{+0.002} \, M_\odot$, $R_2 = 0.218  _{-0.011}^{+0.007} \, R_\odot$, and orbital period of $\sim 4.1$ days. The M-dwarf binary HATS551-027 (LP 837-20) was identified as an eclipsing binary by the HATSouth survey, and characterised by a series of high precision photometric observations of the eclipse events, and spectroscopic determinations of the atmospheric parameters and radial velocity orbits. HATS551-027 is one of few systems with both stellar components lying in the fully-convective regime of very low mass stars, and can serve as a test for stellar interior models. The radius of HATS551-027A is consistent with models to $1\sigma$, whilst HATS551-027B is inflated by 9\% at $2\sigma$ significance. We measure the effective temperatures for the two stellar components to be $T_\text{eff,1} = 3190\pm100$K and $T_\text{eff,2} = 2990\pm110$K, both are slightly cooler than theoretical models predict, but consistent with other M-dwarfs of similar masses that have previously been studied. We also measure significant \ha emission from both components of the binary system, and discuss this in the context of the correlation between stellar activity and the discrepancies between the observed and model temperatures. 
\end{abstract}

\begin{keywords}
(stars:) binaries: eclipsing ; stars: individual:HATS551-027
\end{keywords}

\section{Introduction}
\label{sec:introduction}

Double-lined M-dwarf eclipsing binaries are natural laboratories for stellar astrophysics, providing the most precise, model independent, measurements for the fundamental stellar parameters of low mass stars. They have been used extensively to test our theoretical understanding of stellar interiors \citep[e.g.][]{1995ApJ...451L..29C,2002ApJ...567.1140T}. Very low mass stars (VLMS), with masses below $0.35\,M_\odot$, are thought to have fully convective, adiabatic interiors \citep[e.g.][]{1997A&amp;A...327.1039C,2000ARA&amp;A..38..337C}. Models have been unable to match the observations of VLMS, with measured temperatures cooler and radii larger than models predict  \citep[e.g.][]{2002ApJ...567.1140T,2006Ap&amp;SS.304...89R,2010A&amp;ARv..18...67T,2012ApJ...757...42F,2013ApJ...776...87S,2013AN....334....4T}.

Double-lined M-dwarf binaries in the VLMS mass regime, where the stellar parameters have been accurately determined, are rare. The CM Draconis system is the most well studied, with masses and radii determined to better than 1\% precision \citep{1977ApJ...218..444L,1996ApJ...456..356M,2009ApJ...691.1400M}. The radius discrepancy between models and observations have been well documented for CM Draconis at the 2-10\% level \citep{2002ApJ...567.1140T,2012MNRAS.421.3084M,2013ApJ...776...87S,2014ApJ...789...53F}, depending on the treatment of metallicity, mixing length, and magnetic inhibition in the models. More recently, the systems KOI-126, Kepler-16, LSPM J1112+7626, and WTS 19g-4-02069 \citep{2011Sci...331..562C,2011Sci...333.1602D,2011ApJ...742..123I,2013MNRAS.tmp.1072N} have been reported, with masses and radii of the low mass stellar components measured to better than 2\% precision. Of these systems, only the radii of KOI-126B and C were found to be in agreement with the models of slightly super-solar metallicity, while the other VLMSs are inflated compared to the models \citep{2011ApJ...740L..25F,2012MNRAS.422.2255S,2014ApJ...789...53F,2013MNRAS.tmp.1072N}. 

A number of ideas have been put forward to explain the radius and temperature discrepancies between observations and the models. Increased metallicity and missing opacity may account for the larger measured radii of some systems \citep{2006ApJ...644..475B,2011ApJ...736...47B}. In particular, the KOI-126 B and C companions can be well modelled by assuming a metallicity of \feh=+0.15 \citep{2011ApJ...740L..25F}. However, \citet{2007ApJ...660..732L} found that a metallicity -- radius relationship exists for single stars, but not for binaries. \citet{2012MNRAS.422.2255S} suggested increased metallicity cannot fully account for the model discrepancy for the CM Draconis binary. \citet{2014MNRAS.437.2831Z} found no correlation between metallicity and the measured radii of the VLMS sample, suggesting that the effect of metallicity variation on radius is smaller than the present measurement uncertainties. 

The spin-up of M-dwarfs in binary systems, leading to higher magnetic activities, has also been put forward to resolve the discrepancies \citep[e.g.][]{2007ApJ...660..732L,2007A&amp;A...472L..17C}. \citet{2007A&amp;A...472L..17C} suggested that strong magnetic fields can inhibit convection and heat flow within the stars; increased spot coverage from higher magnetic activities will also modify the evolution of active low mass stars, leading to different radii than inactive stellar models. This is supported by the empirical correlation between X-ray and \ha activity levels and M-dwarf radii \citep{2007ApJ...660..732L,2008A&amp;A...478..507M,2012ApJ...756...47S}. Models of the CM Draconis system by \citet{2012MNRAS.421.3084M,2014ApJ...787...70M} found that magnetic inhibition of convection can explain the radius discrepancy. However, \citet{2014ApJ...789...53F} modelled the Kepler-16 and CM Draconis systems, and found it difficult to sustain the strong magnetic fields required to inhibit convection and inflate the stars to the observed radii.

Recently, accurate M-dwarf parameters have also become important for the characterisation of the planets they host. The larger mass, radius, and luminosity ratios between M-dwarf hosts and planets make them enticing targets for discovery and follow-up characterisation observations. The discrepancies in theoretical modelling of M-dwarfs resulted in a need for empirical relationships, derived from well characterised M-dwarf measurements, in order to characterise the star and planet systems \citep[e.g.][]{2011ApJ...730...79J,2012AJ....143..111J,2014arXiv1408.1758H}.

In this paper, we present the discovery and characterisation of a new eclipsing, double-lined, M-dwarf binary, HATS551-027 (LP 837-20), with both stellar components firmly within the fully convective mass regime of VLMSs. HATS551-027 has previously been identified as a high proper motion \citep{1979nlcs.book.....L,2003ApJ...582.1011S} M4.5 dwarf \citep{2003AJ....126.3007R} from astrometric and low resolution spectroscopic surveys of NLTT high proper motion stars. As part of our effort to characterise the VLMS population \citep{2014MNRAS.437.2831Z}, the HATSouth survey \citep{2013PASP..125..154B} identified and characterised HATS551-027 as a low mass eclipsing binary. We derive precise fundamental properties of the system via the discovery light curves and a series of photometric eclipse follow-up observations and spectroscopic orbit measurements, described in Sections~\ref{sec:observations} and \ref{sec:sys-param}. The HATS551-027 system is then discussed in the context of other VLMS binaries in Section~\ref{sec:discussion}.

\section{Observations and Analysis}
\label{sec:observations}

\subsection{HATSouth Photometric Detection}
\label{sec:hats-phot-detect}

\begin{table*}
  \centering
  \caption{Summary of photometric observations}
  \label{tab:photometric_obs}

  \begin{tabular}{llrrr}
    \hline\hline
    Facility & Dates & Number of Images & Exposure Time (s) & Filter\\
    \hline
    HATSouth Network & 2009 Sep -- 2010 Sep & 16622 & 240 & $r'$\\
    FTS / Merope & 2012 Dec 12 & 124 & 60 & $i'$\\
    Swope / SITe3 & 2013 Feb 26 & 78 & 30 & $i'$\\
    FTS / Merope & 2013 Mar 20 & 114 & 60 & $i'$\\
    \hline
  \end{tabular}
\end{table*}

\input{data/lc.tex}

\begin{figure*}
  \centering
  \includegraphics[width=16cm]{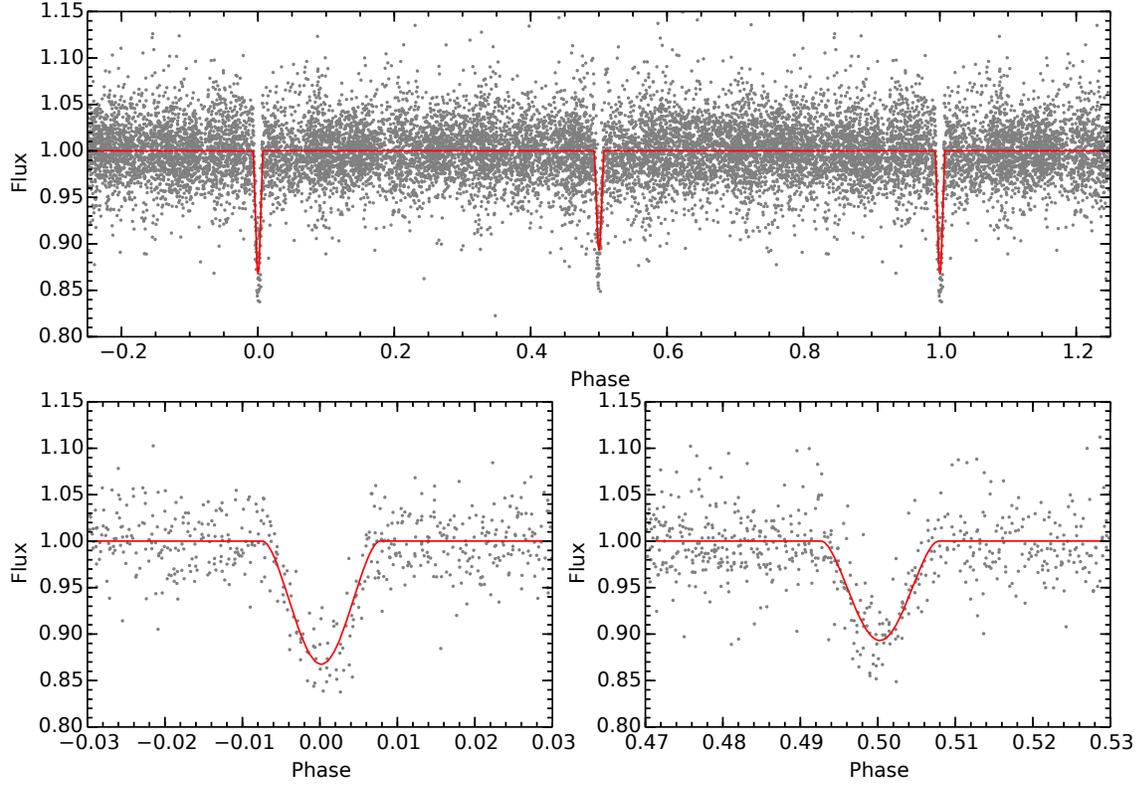}
  \caption{HATSouth discovery light curves of HATS551-027. The observations are plotted in grey, best fit model from Section~\ref{sec:glob-modell-syst} plotted in red. Close-ups of the primary and secondary eclipses are plotted in the lower panels. }
  \label{fig:hslc}
\end{figure*}

The eclipses of HATS551-027 were first identified by observations from the HATSouth survey \citep{2013PASP..125..154B}. HATSouth is a global network of identical, fully robotic telescopes, providing continuous monitoring of selected 128\,deg$^2$ fields of the southern sky. A total of 16622 observations of HATS551-027 were obtained from HATSouth units HS-1, HS-2 in Chile, HS-3, HS-4 in Namibia, and HS-6 in Australia from September 2009 to September 2010. Each HATSouth unit includes four 0.18\,m f/2.8 Takahasi astrographs each with a Apogee 4K$\times$4K U16M Alta CCD camera. The images have a field of view of $4\times4^\circ$, with a pixel scale of 3.7\,''\,pixel$^{-1}$. The observations are performed in the Sloan-$r'$ band, at 4 minute cadence. HATSouth photometry is detrended using the External Parameter Decorrelation \citep[EPD,][]{2007ApJ...670..826B} and Trend Filtering Algorithm \citep[TFA,][]{2005MNRAS.356..557K} techniques. The transit candidates search is perform with the Box-fitting Least Squares \citep[BLS][]{2002A&amp;A...391..369K} analysis. Details of the HATSouth reduction and analysis processes can be found in \citet{2013PASP..125..154B} and \citet{2013AJ....145....5P}. The HATSouth discovery light curve for HATS551-027 is displayed in Figure~\ref{fig:hslc}, the photometric data can be found in Table~\ref{tab:phot_table}.

\subsection{Photometric Follow-up}
\label{sec:phot-observ}

\begin{figure*}
  \centering
  \includegraphics[width=16cm]{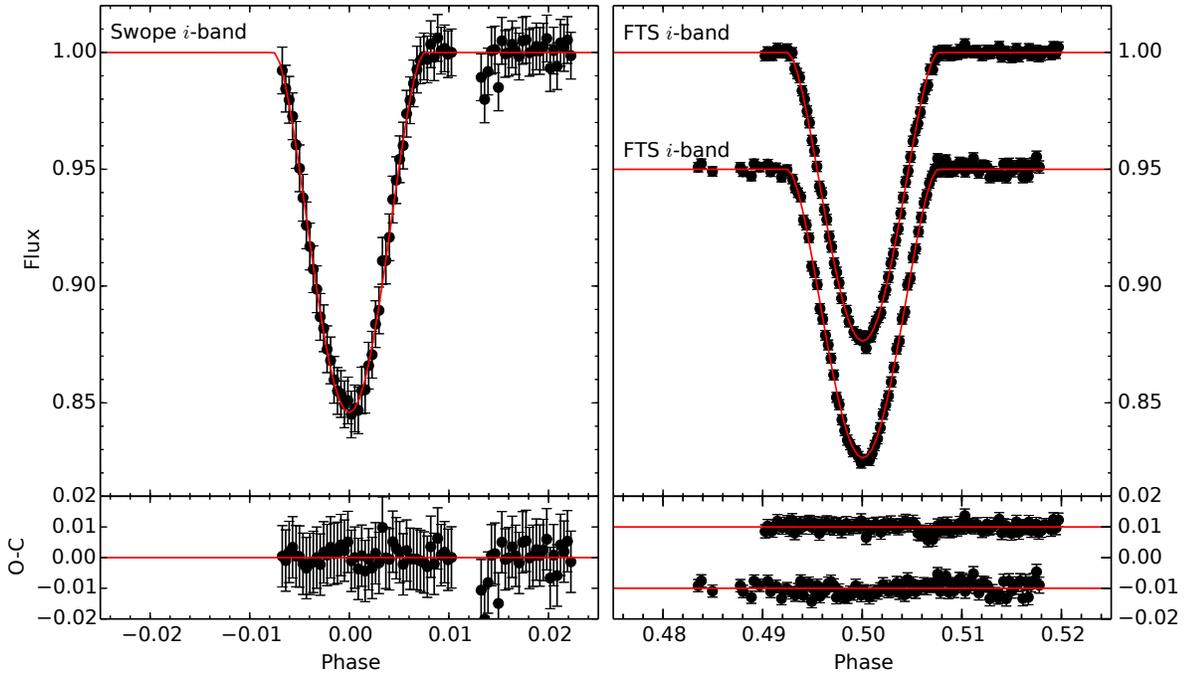}
  \caption{Follow-up photometric observations for the primary (left panel) and secondary eclipses (right panel) of HATS551-027. The observations are plotted in black, best fit model from Section~\ref{sec:glob-modell-syst} plotted in red. The uncertainties shown have been scaled up to force a reduced $\chi^2=1$ where necessary.}
  \label{fig:fulc}
\end{figure*}

We performed photometric followup of the primary and secondary eclipses of HATS551-027. The observations are described below, summarised in Table~\ref{tab:photometric_obs}, with the light curves plotted in Figure~\ref{fig:fulc} and found in Table~\ref{tab:phot_table}. The specific set of reference stars and their magnitudes for differential photometry are listed in Table~\ref{tab:followup_refstars}. 

\begin{table}
  \centering
  \caption{Summary of reference stars for follow-up photometry}
  \label{tab:followup_refstars}

  \begin{tabular}{rllr}
    \hline\hline
    Reference star & RA & DEC & $I$ mag \\
    \hline
    \multicolumn{4}{l}{\emph{FTS 2012-12-12 / 2013-03-20}}\\
    1 & +05:44:56.682 & -24:57:43.844 & 12.1 \\
    2 & +05:44:48.663 & -24:56:55.899 & 14.9 \\
    3 & +05:45:03.111 & -24:56:18.213 & 15.0 \\
    4 & +05:44:59.082 & -24:54:44.068 & 15.7 \\
    5 & +05:45:02.925 & -24:54:28.156 & 15.8 \\
    6 & +05:44:56.505 & -24:54:12.312 & 16.3 \\
    7 & +05:44:49.306 & -24:56:45.252 & 16.2 \\
    8 & +05:45:06.857 & -24:55:22.837 & 16.6 \\
    9 & +05:45:03.468 & -24:56:38.683 & 17.4 \\
    10 & +05:44:51.649 & -24:54:24.864 & 17.0 \\
    11 & +05:45:02.931 & -24:55:32.118 & 15.3 \\
    &&&\\
    \multicolumn{4}{l}{\emph{Swope 2013-02-26}}\\
    1 & +05:45:29.156 & -24:49:56.572 & 12.7 \\
    2 & +05:45:23.524 & -24:51:07.124 & 13.0 \\
    3 & +05:44:32.656 & -24:54:16.506 & 13.4 \\
    4 & +05:44:30.314 & -24:54:57.665 & 13.5 \\
    5 & +05:45:10.934 & -25:02:41.212 & 13.2 \\
    \hline
  \end{tabular}
\end{table}

\subsubsection{Faulkes Telescope South 2\,m / Merope}
\label{sec:phot-foll-up}

Two secondary eclipses of HATS551-027 were observed by the Merope camera on 2\,m Faulkes Telescope South (FTS), at Siding Spring Observatory, on 2012 December 12 and 2013 March 20. FTS/Merope has a field of view of $4.7\times4.7'$, and a pixel scale of $0.139"\,\text{pixel}^{-1}$, read-out using $2\times2$ binning mode. The observations were obtained in the $i$-band, with exposure times of 60 seconds. The telescope was defocused to reduce the effect of intra- and inter-pixel variations on the final light curve, and to avoid saturation of target and reference stars. A total of 124 observations were obtained on 2012 December 12, and 114 on 2013 March 20. 

Bias and flat field reduced images, automatically generated by the LCOGT reduction pipeline, were used for the photometric extraction. The raw light curves for the target and reference stars were extracted using Source Extractor \citep{1996A&amp;AS..117..393B},  via multiple fixed circular apertures, and after an interpolated background map was subtracted. Eleven reference stars of magnitudes $I=12.1$--17.4 were used for the relative photometry. The best aperture was chosen to minimise the out-of-transit light curve scatter. A second order polynomial was fitted to the out-of-transit regions of the light curves to remove small, long-duration variations, likely due to changing airmass, before model fitting was performed. 

\subsubsection{Swope 1\,m / SITe3}
\label{sec:swope-1msite3}

A near-complete primary eclipse of HATS551-027 was observed by the SITe\#3 camera on the Swope 1\,m telescope at Las Campanas Observatory, Chile, on 2013 February 26. Swope/SITe3 has a field of view of $14.8\times22.8'$, a pixel scale of 0.435''\,pixel$^{-1}$, and detector size of $2048\times3150$ pixels. The observations were performed in the $i$-band, with 30s exposures. A total of 78 exposures were obtained. Photometric extraction were performed on the reduced frames using the FITSH package \citep{2012MNRAS.421.1825P}. Five reference stars of magnitudes $I=12.7$--13.5 were used for the relative photometry. We note that no pre-ingress baseline was recorded, which may have an adverse effect on the derived timings and associated uncertainties. 

\subsection{Spectroscopic Follow-up}
\label{sec:spectr-observ}

\begin{table*}
  \centering
  \caption{Summary of spectroscopic observations}
  \label{tab:spec_obs}

  \begin{tabular}{llrrr}
    \hline\hline
    Facility & Date Range & Number of Observations & Resolution & S/N$^a$\\
    \hline
    ANU 2.3m / WiFeS & 2012 Sep 8 & 1 & 3000 & 95\\
    ANU 2.3m / WiFeS & 2012 Sep 8 -- 2014 Aug 14 & 16 & 7000 ($R$ band) & 20\\
    MPG/ESO 2.2m / FEROS & 2013 Mar 24 -- 27 & 4 & 48000 & 35\\
    Magellan 6.5m / PFS & 2013 Feb 21 & 4 & 76000 & 90 \\
    ANU 2.3m / WiFeS & 2014 Sep 6 -- 2014 Oct 10 & 3 & 7000 ($I$ Band) & 60\\
    \hline
  \end{tabular}
\begin{flushleft} 
$^a$ Average signal-to-noise ratio per resolution element, at approximate peak. 
\end{flushleft}
\end{table*}

\begin{figure*}
  \centering
  \includegraphics[width=14cm]{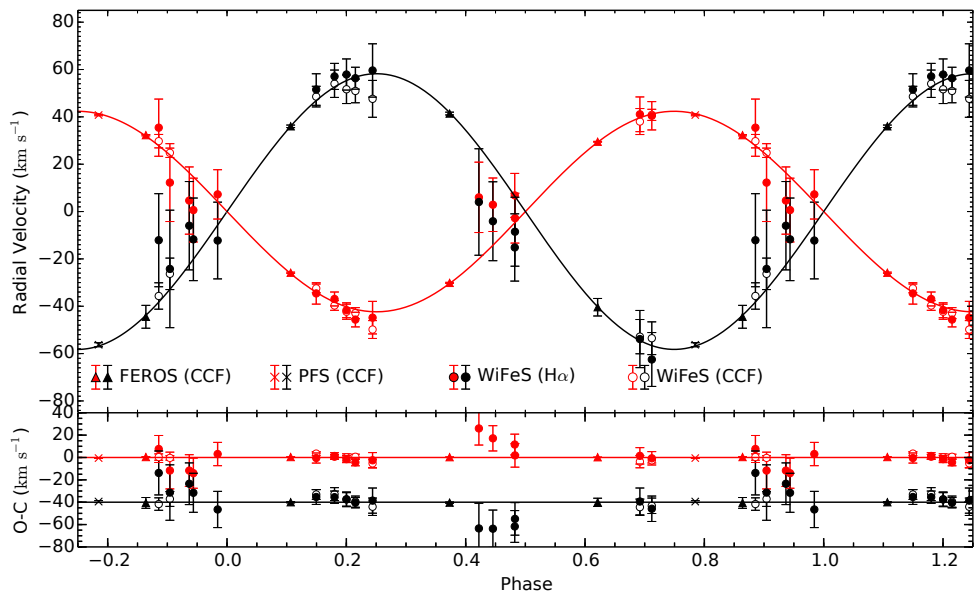}
  \caption{The radial velocity orbit of HATS551-027A (red) and B (black). The best fit models from Section~\ref{sec:glob-modell-syst} are plotted by the red and black curves for components A and B respectively.}
  \label{fig:rv}
\end{figure*}

We performed a series of spectroscopic follow-up observations at low, medium, and high resolution, to determine the stellar properties and orbit of the system. The observations are summarised in Table~\ref{tab:spec_obs}, and the radial velocities are plotted in Figure~\ref{fig:rv} and presented in Table~\ref{tab:rv_table}. 

\input{data/rv.tex}

\subsubsection{ANU 2.3\,m / WiFeS}
\label{sec:anu-2.3m}

Spectroscopic observations at low and medium resolutions were performed using the Wide Field Spectrograph \citep[WiFeS,][]{2007Ap&amp;SS.310..255D} on the ANU 2.3m telescope, located at Siding Spring Observatory, Australia. WiFeS is an image slicer integral field spectrograph, with slitlets of 1'' width in spatial coverage. The WiFeS observations provided the spectroscopic classification and radial velocities (to the $\sim 2$\kms level) of HATS551-027. 

For spectral classifications, an observation at low resolution of $\lambda/\Delta \lambda \equiv R = 3000$ was obtained using the B3000 and R3000 gratings on the blue and red arms of the spectrograph, with the RT560 dichroic, providing a flux calibrated spectrum of the object over 3500--9000\,\AA. For the $R=3000$ observations, the stellar flux is summed over the three slitlets together to produce the final spectrum. Flux calibrations are performed according to \citet{1999PASP..111.1426B}, using flux standard stars from \citet{1992PASP..104..533H} and \citet{1999PASP..111.1426B} taken on the same night. The observational techniques, data reduction and flux calibration processes for the WiFeS low resolution observations are fully described in \citet{2013AJ....146..113B}. 

To measure the radial velocity orbit, we then obtained 16 medium resolution $(R=7000)$ observations from September 2012 to August 2014. The observations were performed using the R7000 grating and the RT480 dichroic, giving a velocity resolution of 21.6\,\kms\,pixel$^{-1}$ over the wavelength range 5200--7000\,\AA, corresponding to the photometric $R$ band. The wavelength calibrations were obtained from bracketing Fe-Ne-Ar arc lamp exposures. In addition, a first order correction to the wavelength calibration is applied using the telluric Oxygen B lines over 6882--6906\,\AA. For radial velocity measurements, we reduce the spectra from each WiFeS slitlet separately, such that three reduced spectra from three slitlets are produced per exposure. The reduction process is described in detail in \citet{2013AJ....146..113B}. We adopted the \ha-derived velocities for analysis in Section~\ref{sec:glob-modell-syst}.

\begin{figure}
  \centering
  \includegraphics[width=9cm]{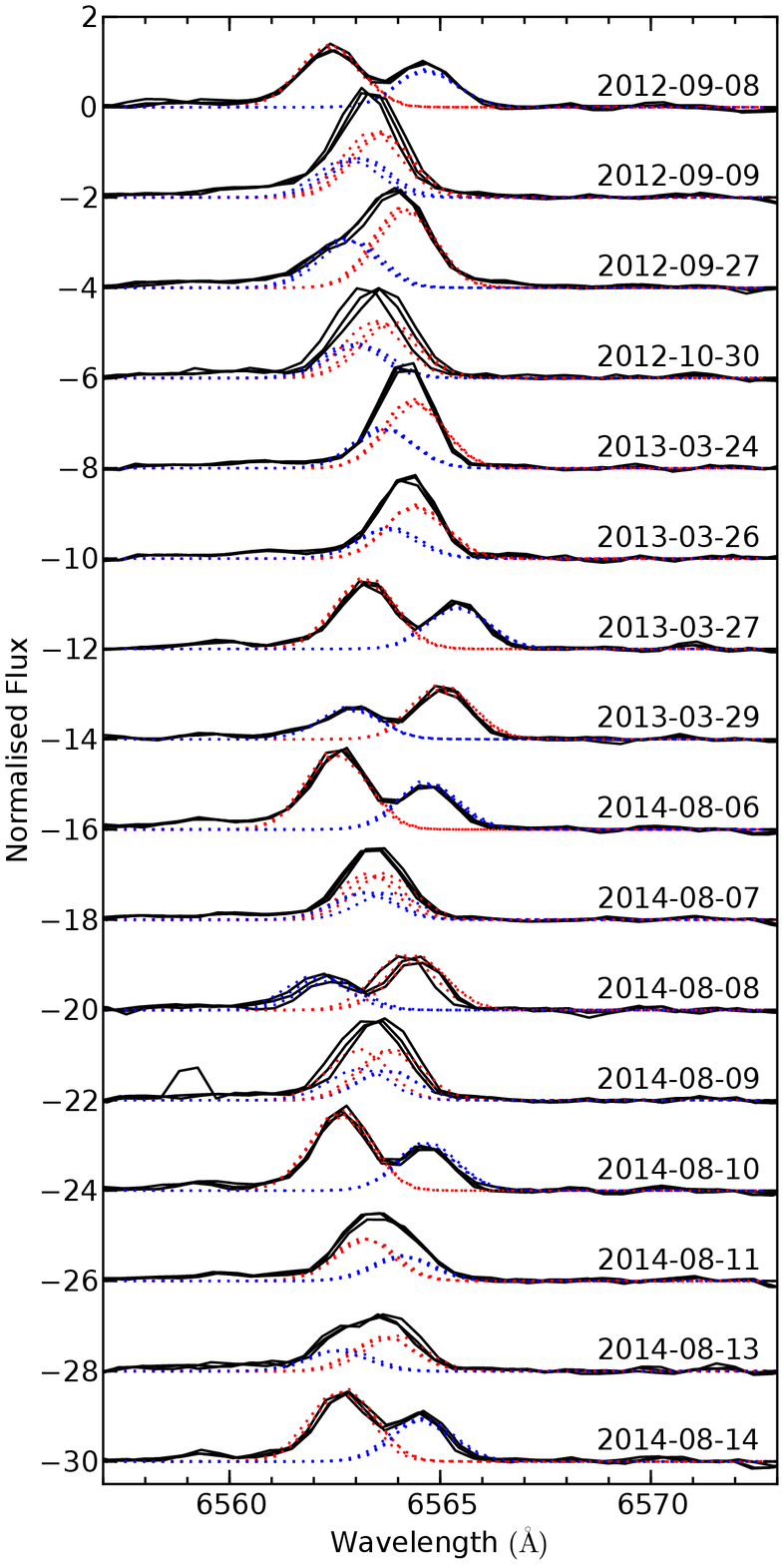}
  \caption{The \ha emission feature from the WiFeS $R=7000$ observations. For each exposure, the spectra from the three slitlets are plotted individually, in black. The best fit profiles are plotted in red for the primary component, and blue for the secondary component. Each exposure is arbitrarily offset in flux by 2 for clarity.}
  \label{fig:Ha_fit}
\end{figure}

Radial velocities of both components of the binary were first measured by fitting the \ha emission feature, which exhibits clear and distinct signals from both stellar companions. For each spectrum, we fit the \ha feature with two Gaussian components. The best fit values are obtained via a down-hill simplex minimisation using the \emph{Python} \emph{fmin} function in \emph{scipy} package, and the errors explored via an MCMC analysis using the \emph{emcee} affine invariant ensemble sampler \citep{2013PASP..125..306F}. The final velocity error from each exposure is a quadrature combination of the mean centroid errors from the MCMC fit and the scatter in velocity from three slitlets. The \ha emission features and Gaussian fits for each observation are presented in Figure~\ref{fig:Ha_fit}. The observations from each night are plotted by three black solid lines, for the spectra from the three slitlets. The best fit Gaussians for each spectrum are also plotted by the dotted line, in red for the primary component, blue for the secondary component. The velocities from WiFeS are presented in Figure~\ref{fig:rv} and Table ~\ref{tab:rv_table}.

We also tried cross correlations between the target spectra and an M4V template from a WiFeS observation of the standard star LP 816-60. The radial velocity standard cross correlations provided blended cross correlation functions (CCFs), which did not resolve the two spectral components as well as the \ha feature. We derived radial velocities from selected WiFeS R7000 observations taken at phase quadrature by simultaneously fitting a double Gaussian to the CCF profile, presented in Figure~\ref{fig:rv} and Table~\ref{tab:rv_table}. We also tried cross correlating against spectral templates from the BT-Settl model atmospheres \citep{2012EAS....57....3A}, but we could not resolve the two spectral components.

The WiFeS cross correlations were also used to measure light ratio between the two stellar components $(L_2/L_1)$. We fit a double Gaussian to the CCFs from the spectra, and derived a light ratio of $0.53\pm0.06$ from the R7000 grating, equivalent of the photometric $R$ band. We also obtained three additional exposure over the wavelength range 6800--9100\,\AA\, at $R=7000$, using I7000 grating and the RT615 dichroic, corresponding to the photometric $I$ band. This exposure was used to measure the light ratio between the two components at the wavelength of the follow-up photometric observations. The I7000 observation gave a light ratio of $0.50\pm0.01$, consistent to the $R$ band light ratio.

\subsubsection{MPG 2.2\,m / FEROS}
\label{sec:eso-mpg-2.2m}

Four observations of HATS551-027 were obtained using the fibre-fed FEROS spectrograph on the MPG 2.2m telescope at La Silla Observatory, Chile , over the nights 2013-03-24 to 2013-03-27. FEROS provides a resolution of $R=48000$, descriptions of the FEROS observations and reduction procedures can be found in \citep{2013AJ....145....5P,2014AJ....148...29J,2015arXiv150300062B}. The observations are cross correlated with an M-dwarf template. The two components of the binary are fully separated in the CCF, and the velocities are presented in Figure~\ref{fig:rv} and Table~\ref{tab:rv_table}. We also measure the Gaussian peak heights of the CCFs to estimate the light ratio for the two components of the system. Because of the faintness of the target star, we find that two orders in the wavelength range 6475--6665\,\AA\,  reliably showed the two CCF peaks in all of the exposures. The average and standard deviation CCF peak heights over the 4 exposures was $0.42\pm0.07$. 

\subsubsection{Magellan 6.5\,m / PFS}
\label{sec:magellan--pfs}

Four consecutive observations of HATS551-027 were obtained using the Carnegie Planet Finder Spectrograph \citep[PFS,][]{2010SPIE.7735E..53C} on Magellan II, Las Campanas Observatory, Chile, on the night of 2013-02-21 UT. The observations were taken through the iodine cell, with a $0.5"\times 2.5"$ slit and using $2\times2$ binning in slow read-out mode. 

We use the PFS spectrum to derive the single epoch relative radial velocities and light ratio of the two stellar components. The spectra from the four observations were averaged, and cross correlated against a $T_\text{eff} = 3100\,\text{K}$, $\log g = 5.0$ solar metallicity BT-Settl synthetic spectrum \citep{2012EAS....57....3A}. We used five orders outside of the Iodine absorption region (6114--6574\,\AA) for the cross correlation. The two stellar components are resolved and fully separated in the CCF. We simultaneously fit two Gaussians to the CCF, deriving the light ratios and the velocity centroids. The velocities from the five orders are weight averaged to arrive at the final relative velocity between the two stellar components. The light ratio $L_2/L_1$ is $0.51\pm0.04$, with the error as the standard deviation between the five orders. We also tried 3000\,K and 3200\,K synthetic templates, yielding the same light ratio to within $\pm 0.01$. The light ratio measured from PFS is also consistent to $\sim 1\sigma$ of that measured from the FEROS spectra and WiFeS spectra. We adopt the light ratio measured from the high signal-to-noise PFS observations for our modelling in Section~\ref{sec:glob-modell-syst}.

\section{System parameters}
\label{sec:sys-param}

\subsection{Global modelling of light curves and radial velocities}
\label{sec:glob-modell-syst}

We perform a global modelling of the HATSouth discovery light curves, follow-up photometry, and spectroscopic radial velocities to derive the system parameters for HATS551-027. We model the light curves according to \citet{1972ApJ...174..617N}, using a modified version of the JKTEBOP code \citep{1981AJ.....86..102P,2004MNRAS.351.1277S}. The radial velocities are modelled using Keplerian orbits. 

The free parameters of the fit are period $P$, reference transit time $t_0$, radius ratio $R_2/R_1$, normalised radius sum $(R_1+R_2)/a$, line-of-sight inclination $i$, light ratio $L_2/L_1$, orbital eccentricity parameters $e\cos\omega$ and $e\sin\omega$, and orbital semi-amplitudes $K_1$, $K_2$ for the two components. We assign a Gaussian prior to $L_2/L_1$ \citep[following e.g. ][]{2011ApJ...742..123I} from the PFS light ratio of $0.51\pm0.04$ from Section~\ref{sec:magellan--pfs}. Without the prior, the radius ratio becomes degenerate with the light ratio, and is difficult to constrain. We assume uniform priors for all other free parameters. The quadratic limb darkening coefficients for both components are fixed to that from \citet{2000A&amp;A...363.1081C} using the PHOENIX models, and given in Table~\ref{tab:sys_param}. At each iteration, the modelled mass ratio $(q=K_1/K_2)$ is used to calculate the ellipsoidal variability expected in the light curve. Following \citet{2009ApJ...691.1400M}, we assign fixed reflection albedos for both components of 0.5, and gravity darkening coefficients of 0.2. The instrument zero points for each dataset are not included as free parameters in the global model, but rather fitted for at each iteration independently. We often find the HATSouth discovery light curves to be diluted due to the TFA detrending that was applied, as well as third-light blending from nearby neighbours due to the larger pixel scales of the instrument. The model transit shape is significantly constrained by the follow-up observations, and can be used to correct for the dilution of the HATSouth light curves. At each iteration, we adjust the transit depth of the HATSouth light curves to the tested model by fitting for a third light component \citep[e.g.][]{2014MNRAS.437.2831Z}. We adopt the WiFeS radial velocities derived from the \ha emission line in our analysis, since the two stellar components are better resolved in \ha than in the CCFs.

The best fit parameters and the associated uncertainties are explored via a Markov chain Monte Carlo (MCMC) analysis, using \emph{emcee}. For each MCMC run, we initialise 100 walkers over 2500 steps each. The walkers are initialised with randomised starting positions around their respective expected best fit parameter values. Examination of the MCMC chain shows that convergence is generally reached after 500 -- 1000 steps, depending on the initial conditions. We then re-run the MCMC with walkers initialised around the best fit parameters from the first run. We inflate the per-point measurement uncertainties such that the reduced \chisq for each dataset is at unity with respect to the best fit parameters derived from the first MCMC run. This allows other sources of error unaccounted for in data reduction to be included in the MCMC analysis. We found that only the FTS observations required their measurement uncertainties to be inflate, by $2.4\times$ and $2.1\times$ for the 2012-12-12 and 2013-03-20 observations respectively.

\subsubsection{Accounting for Spots}
\label{sec:accounting-spots}

Late M-dwarfs are known to be active, and the presence of chromospheric spectral emission features of HATS551-027 suggest that it is no exception. A Lomb-Scargle  \citep[LS,][]{1976Ap&amp;SS..39..447L,1982ApJ...263..835S} analysis of the out-of-transit HATSouth light curve residuals from Section~\ref{sec:glob-modell-syst} shows a highest peak at 4.24 days, with peak FWHM of 0.08 days, compared to an orbital period of 4.08 days. Although when we break the light curve into four segments of 4000 points each, we cannot demonstrate the consistency of the peak location.

\begin{figure*}
  \centering
  \includegraphics[width=15cm]{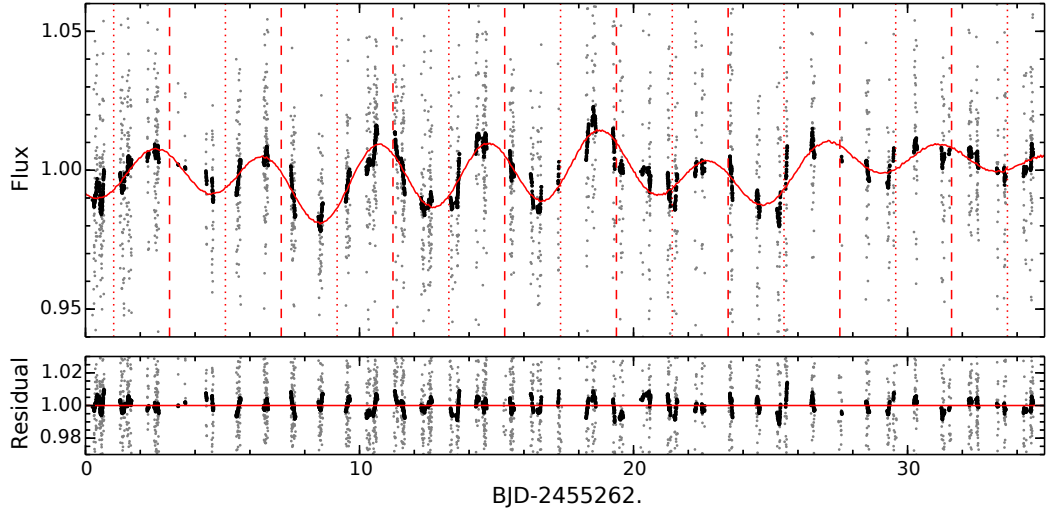}
  \caption{A 35 day sample segment of the HATSouth light curve residuals to the best fit model from Section~\ref{sec:glob-modell-syst}. The data points are plotted in grey, 50 point-width boxcar smoothed function in black, and the cosine filtering model by the red line. The times of transits and eclipses, masked out in the data, are marked by vertical red dashed and dotted lines, respectively. The corrected data are plotted on the bottom panel. }
  \label{fig:HS_cosfilt}
\end{figure*}

Star spot activity affect the observations and model results in several ways. First, they induce periodically modulated out-of-transit variations in the HATSouth light curves. We correct this by modelling the cosine filtered light curve in the MCMC analysis. We model the modulation in the HATSouth light curve residuals using a set of linearly combined cosine functions, following Equation 1 of \citet{2013MNRAS.429.2001H}. The HATSouth light curve residuals, with transits and eclipses masked, are smoothed by a 50 point-width boxcar, and fitted with the cosine filtering model with minimum period of $P/2$ and maximum period of twice the observational baseline. The light curve residuals and cosine filtering model for a segment of the HATSouth observations are shown in Figure~\ref{fig:HS_cosfilt}. The cosine filtering model has a standard deviation of 0.006, and peak-to-peak amplitudes of $\sim 2$\%. For comparison, the modelled out-of-transit variations originating from reflection and ellipsoidal effects is at the $10^{-4}$ level. 

Variations in spot coverage also alters the total flux of the system, leading to variations in the eclipse depth. This affects the parameters derived from the high precision follow-up photometry, such as $R_2/R_1$ and $(R_2+R_1)/a$. Since the follow-up observations were performed $\sim 800$ days after the end of the HATSouth continuous observations, they cannot be corrected by employing spot model fits to the out-of-transit variability \citep[e.g.][]{2009ApJ...691.1400M}.  Instead, we create 100 different sets of the follow-up light curves, with each light curve scaled by a random factor drawn from a Gaussian distribution with $\mu=1.0$ and $\sigma=0.006$ (the standard deviation of the cosine filtering model). We then perform a global analysis detailed in Section~\ref{sec:glob-modell-syst} using the scaled sets of the follow-up light curves. The MCMC chains from each of the 100 separate analyses are combined to provide the final system uncertainties. 

In-transit star spot crossing events modify the shape of the transit light curve, and can be identified from the correlated red-noise in the fit residuals. Whilst we cannot correct for the spot crossing events, we can account for their effects in the uncertainties of the model fit. To account for correlated noise, we treat the follow-up photometry as a Gaussian process \citep[e.g.][]{2012MNRAS.419.2683G}. We form the covariance matrix $\Sigma$, of dimension $N\times N$, where $N$ is the number of data points per follow-up light curve. The elements of the covariance matrix is defined as
\begin{equation}
  \label{eq:covariance_matrix}
  \Sigma_{ij} = A^2 \exp \left( - \frac{(t_i - t_j)^2}{2\tau^2}\right) + \mathbf{\sigma}_i^2 \delta _{ij}\,,
\end{equation}
where $A$ and $\tau$ are free parameters in the MCMC describing the amplitude and timescale of the covariance function, and $\sigma_i$ is the uncorrelated uncertainty for measurement $i$. The log likelihood $\mathcal{L}$ for the residuals $\mathbf{r}$ is calculated as:
\begin{equation}
  \label{eq:loglike}
  \log \mathcal{L} = -\frac{1}{2} \mathbf{r}^T \Sigma^{-1} \mathbf{r} - \frac{1}{2} \log \mid \Sigma \mid - \frac{N}{2} \log (2\pi)\,.
\end{equation}
The Gaussian process modelling provides more realistic uncertainty estimates. The $A$ and $\tau$ factors are presented in Table~\ref{tab:sys_param}

\subsubsection{Results}
\label{sec:results}

% \begin{align}
%   \label{eq:dynam_masses}
%   m_1 &= (1-e^2)^{3/2} \frac{K_2 (K_1+K_2)^2 P}{2 \pi G \sin(i)^3}\\
%   m_2 &= (1-e^2)^{3/2} \frac{K_1 (K_1+K_2)^2 P}{2 \pi G \sin(i)^3}\,.
% \end{align}

The final masses and radii are $M_1 = 0.244 _{-0.003}^{+0.003} \, M_\odot$, $R_1 = 0.261_{-0.009}^{+0.006} \, R_\odot$,  $M_2 = 0.179 _{-0.001}^{+0.002} \, R_\odot$, $R_2 = 0.216  _{-0.011}^{+0.007} \, R_\odot$. The full set of system parameters are presented in Table~\ref{tab:sys_param}. The results presented are the mode of the posterior probability distributions, with the uncertainties representing the 68\% confidence region. 

\begin{figure*}
  \centering
  \includegraphics[width=19cm]{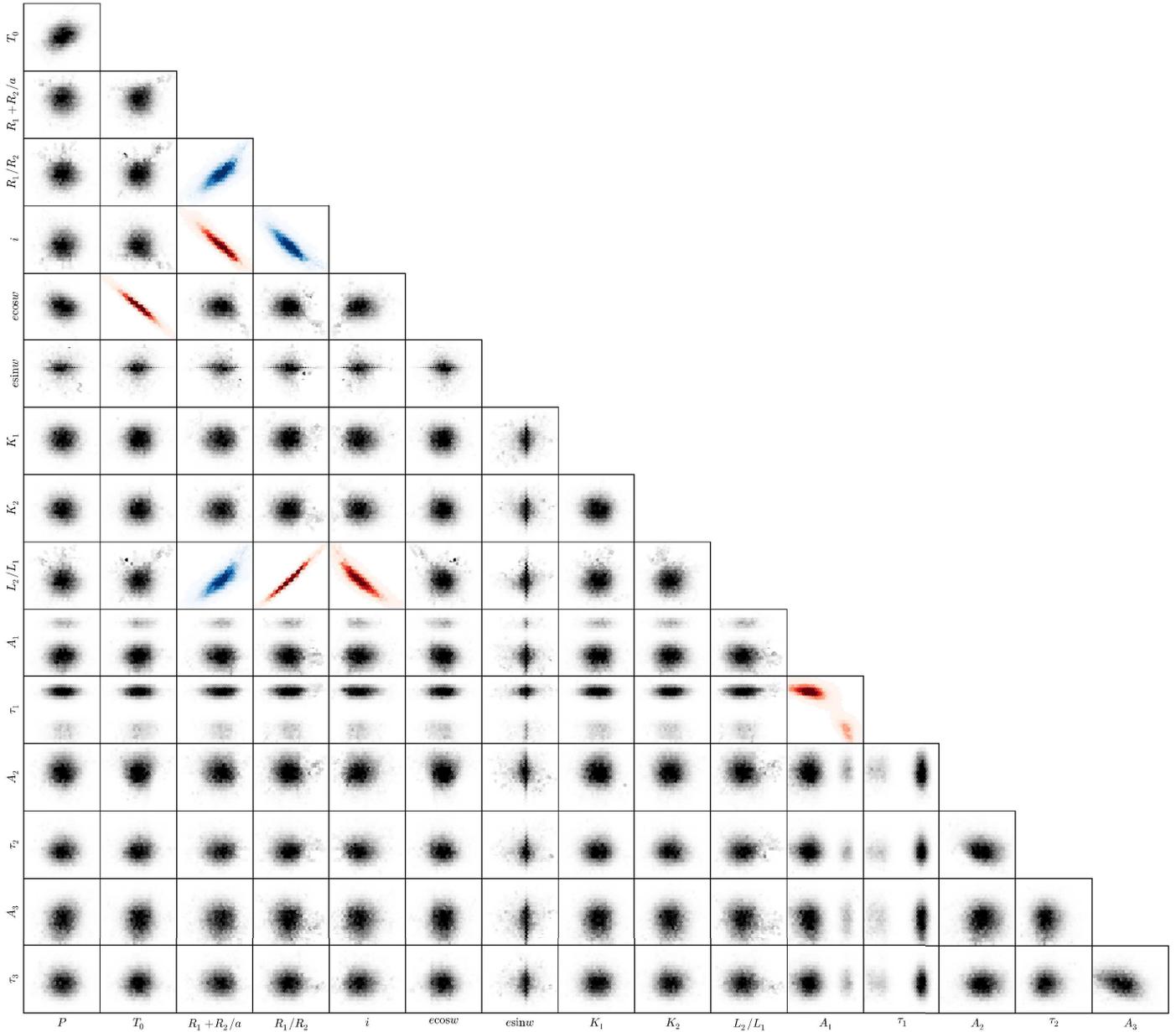}
  \caption{Posterior correlation diagrams for the free parameters in the global fit. Pairs of parameters with Pearson correlation coefficient of $0.5 < R < 0.8$ are coloured in blue, $R > 0.8$ coloured in red.}
  \label{fig:posterior-all}
\end{figure*}

The posterior probability distributions for the free parameters in our global fit are shown in Figure~\ref{fig:posterior-all}. For each pair of parameters, we calculate the Pearson $R$ correlation coefficient to check for degeneracies. We note a significant degeneracy between the light ratio $L_2/L_1$ and the radius ratio $R_2/R_1$ in our results. For systems with partial eclipses, it is difficult to constrain both parameters simultaneously purely from the transit light curves without a prior on the light ratio. We also note a correlation between $(R_1+R_2)/a$ and inclination $i$, since both parameters together determine the transit shape. The time parameters $T_0$ and $e\cos\omega$ are correlated, since the phase of the secondary eclipse, determined by $e\cos\omega$, is influenced by the ephemeris of the primary transit $T_0$.   
We examined the posterior probability distributions for the mass and radius of both stars, and found no correlation between the derived mass vs radius distribution.

\input{system_parameters.tex}

\subsubsection{Testing the robustness of the results}
\label{sec:fit_result_robustness}

To test the dependence of our results against various assumptions and datasets, we re-ran the full MCMC analysis with different subsets of data and initial parameters. For each test, we describe the changes to the free parameters, assumptions, and datasets used, and report the major differences of the derived results to those reported in Table~\ref{tab:sys_param}. While these tests provide a useful guide for gauging the robustness of our results, the uncertainty estimates we report are likely still optimistic, as they are largely dependent on the per-point measurement uncertainties. Only repeated velocity measurements and photometric follow-up observations will provide the most robust stellar properties for HATS551-027.

\begin{itemize}

\item We set limb darkening coefficients as free parameters, but found that we can only place very weak constraints on the coefficients from the fitting. In $R$ band, component A has best fit limb darkening coefficients of $u_1=0.5_{-0.3}^{+0.5}$, $u_2=0.3_{-0.4}^{+0.4}$, component B of $u_1=0.7_{-0.3}^{+0.5}$, $u_2=0.0_{-0.4}^{+0.3}$, and in $I$ band, component A of $u_1=0.7_{-0.5}^{+0.9}$, $u_2=0.1_{-0.5}^{+0.7}$, component B of $u_1=0.7_{-0.5}^{+0.9}$, $u_2=0.1_{-0.6}^{+0.6}$. We found the limb darkening parameters were largely orthogonal to the other parameters. The resulting best fit parameter values did not differ significantly between the limb darkening free and fixed analyses, and the radii and mass uncertainties were not noticeably inflated by freeing the limb darkening parameters.

\item To test the effect of the single epoch of PFS velocities on the resulting derived masses, we refit the observations without the PFS velocities. The resulting orbit semi-amplitudes were $K_1 = 42.4_{-0.2}^{+0.1}\,\text{km}\,\text{s}^{-1}$ and $K_2 = 58.0_{-0.4}^{+0.6}\,\text{km}\,\text{s}^{-1}$, yielding masses of $M_1 = 0.249_{-0.005}^{+0.004}\,\text{M}_\odot$ and $M_2 = 0.181_{-0.002}^{+0.002}\,\text{M}_\odot$. The derived properties are consistent with the global fit results to within errors.

\item To test if the \ha derived velocities, which reflect velocities of the chromosphere, are representative of the photospheric velocities, we fit the system using only WiFeS CCF derived velocities. We derived velocity semi-amplitudes of $K_1 = 44.1_{-1.0}^{+0.9}\,\text{km}\,\text{s}^{-1}$ and $K_2 = 53.2_{-2.2}^{+2.2}\,\text{km}\,\text{s}^{-1}$, yielding masses of $M_1 = 0.21_{-0.02}^{+0.02}\,\text{M}_\odot$ and $M_2 = 0.18_{-0.01}^{+0.01}\,\text{M}_\odot$. In comparison, using only the WiFeS \ha derived velocities, we derive semi-amplitudes of $K_1 = 43.3_{-1.4}^{+1.4}\,\text{km}\,\text{s}^{-1}$ and $K_2 = 59.7_{-2.6}^{+2.0}\,\text{km}\,\text{s}^{-1}$, yielding masses of $M_1 = 0.26_{-0.02}^{+0.02}\,\text{M}_\odot$ and $M_2 = 0.19_{-0.01}^{+0.01}\,\text{M}_\odot$. The discrepancy between the $K_2$ values is likely due to the difficulties in resolving and fitting the secondary CCF and \ha component using the lower resolution spectra. The derived semi-amplitudes and masses are consistent to the global fit results to within $2\sigma$. 

\item To test the dependence of the results on the four FEROS measurements and associated uncertainties, we fit the dataset with only the FEROS velocities and derive velocity semi-amplitudes of $K_1 = 42.4_{-0.3}^{+0.2}\,\text{km}\,\text{s}^{-1}$ and $K_2 = 57.8_{-0.8}^{+1.3}\,\text{km}\,\text{s}^{-1}$, yielding masses of $M_1 = 0.247_{-0.009}^{+0.010}\,\text{M}_\odot$ and $M_2 = 0.182_{-0.005}^{+0.004}\,\text{M}_\odot$. The masses determined are consistent, but have three times larger uncertainties from the global fit. To check the dependence of our derived masses against the measurement precision, we inflate the FEROS uncertainties by $2\times$, and re-fit using only the inflated FEROS velocities, deriving semi-amplitudes of $K_1 = 42.4_{-0.5}^{+0.5}\,\text{km}\,\text{s}^{-1}$ and $K_2 = 58.0_{-2.0}^{+1.8}\,\text{km}\,\text{s}^{-1}$, yielding masses of $M_1 = 0.248_{-0.018}^{+0.017}\,\text{M}_\odot$ and $M_2 = 0.181_{-0.009}^{+0.007}\,\text{M}_\odot$. Inflating the velocity measurement uncertainty by two times leads to a $\sim 2$ times inflation of the derived mass uncertainties.

\item In Section~\ref{sec:accounting-spots}, we tried to account for spot-induced uncertainties by 1) Monte Carlo of out-of-transit baseline flux and 2) addition of Gaussian processes modelling. To test their effects, we removed the two treatments and re-fit the data. The derived radius ratio is $R_2/R_1=0.82_{-0.04}^{+0.03}$, light ratio $L_2/L_1=0.53_{-0.05}^{+0.04}$, resulting stellar radii of $R_1 = 0.260_{-0.004}^{+0.006}\,R_\odot$ and $R_2=0.214_{-0.006}^{+0.006}\,R_\odot$. Note the significant reduction in the estimated uncertainties ($\sim 60$\%) without accounting for spot activity.

\item To test the dependence of the results on the systematic errors associated with each FTS secondary eclipse observation, and the polynomial normalisation applied to the light curves, we fit the observations using each FTS secondary eclipse observation independently. Using the FTS 2012-12-12 light curves for the secondary eclipse, we derive a radius ratio of $R_2/R_1=0.84_{-0.07}^{+0.06}$, light ratio $L_2/L_1=0.55_{-0.09}^{+0.09}$, resulting stellar radii of $R_1 = 0.259_{-0.007}^{+0.009}\,R_\odot$ and $R_2=0.22_{-0.01}^{+0.01}\,R_\odot$. Using the FTS 2013-03-20 light curves for the secondary eclipse, we derive a radius ratio of $R_2/R_1=0.81_{-0.07}^{+0.07}$, light ratio $L_2/L_1=0.52_{-0.10}^{+0.09}$, resulting stellar radii of $R_1 = 0.262_{-0.008}^{+0.010}\,R_\odot$ and $R_2=0.22_{-0.02}^{+0.01}\,R_\odot$. In each case, the derived values are consistent, with the uncertainties for the radius and light ratio parameters increased by $\sim 30$\,\%.

\item In the Swope light curve, four measurements at phase $\sim 0.15$ were consistently below the out-of-transit median. We masked out the outlying measurements and performed the refit, deriving a radius ratio of $R_2/R_1=0.84_{-0.05}^{+0.05}$, light ratio $L_2/L_1=0.55_{-0.07}^{+0.07}$, resulting stellar radii of $R_1 = 0.261_{-0.007}^{+0.006}\,R_\odot$ and $R_2=0.22_{-0.01}^{+0.01}\,R_\odot$. The differences are negligible compared to the global fit. 

\item We note that the follow-up photometric observations were all obtained in the $i$ band. Single-band follow-up observations can bias the derived radius ratio parameters if the light ratio is band-dependent. For comparison, we fit the observations using the $r$-band HATSouth observations only, without the photometric follow-up datasets, we derive $T_0=2456374.017_{-0.002}^{+0.002}$ (consistent, but 10 times worse in precision), $(R_1+R_2)/a=0.060_{-0.002}^{+0.002}$ (consistent but 4 times worse in precision), $R_2/R_1=0.83_{-0.05}^{+0.04}$ (consistent and equal precision), light ratio of $L_2/L_1=0.52_{-0.07}^{+0.04}$ (consistent and equal precision), and component radii of $R_1 = 0.267_{-0.012}^{+0.009}$ and $R_2=0.22_{-0.01}^{+0.01}$ (a larger $R_1$, but still consistent within 1$\sigma$). 

\item The light ratio $L_2/L_1$ in our global fit is tightly constrained by a Gaussian prior of mean 0.51 and standard deviation 0.04, determined by the analyses of the PFS spectra. To understand the effect of this prior on our resulting parameters, we modify to the Gaussian prior to have a mean 0.51 and standard deviation 0.08 ($2\times$ broader). The resulting fit has a $R_2/R_1=0.88_{-0.08}^{+0.09}$, light ratio $L_2/L_1=0.64_{-0.14}^{+0.10}$, resulting stellar radii of $R_1 = 0.251_{-0.008}^{+0.012}\,R_\odot$ and $R_2=0.23_{-0.02}^{+0.01}\,R_\odot$. The results remain consistent with $1\sigma$, with the radius uncertainties increased by a factor of $\sim 1.5$.

\end{itemize}

\subsection{Spectral Classification}
\label{sec:spectr-class-from}

\subsubsection{Effective Temperature}
\label{sec:temperature}

We use the WiFeS low resolution spectrum, which has the largest spectral coverage and highest S/N, to constrain the effective temperature of HATS551-027. Whilst the spectrum of HATS551-027 is a composite, at low resolution we cannot resolve the two binary components by their velocity differences. We first treat the spectral classification as if HATS551-027 is a single star. 

The flux calibrated WiFeS $R=3000$ observation is matched to synthetic spectra from BT-Settl models \citep{2012EAS....57....3A} with \citet{2009ARA&amp;A..47..481A} abundances. We fix \logg to 5.0 \citep[derived from Section~\ref{sec:results} to be 4.99 and 5.01 for stars A and B, and consistent with stellar isochrone models, e.g. ][]{2008ApJS..178...89D} and [M/H] to 0.0, and minimise the \chisq fit by varying \teff at steps of 100\,K, interpolating the \teff --\chisq relationship with a B-spline. 

First, to understand the uncertainties in our temperature classification, we obtained WiFeS spectra of five M-dwarfs with temperatures measured from interferometry observations. The WiFeS BT-Settl model fit temperatures and the interferometry temperatures are shown on Table~\ref{tab:wifes_mdwarf_char}. We find an error of 85\,K encompasses 68\% of the deviations between the WiFeS measurements and literature values, weighted by the literature uncertainties. We find an error of 150\,K when we use models employing the \citet{2011SoPh..268..255C} abundances.  We also tested the standard stars against models with metallicities of [M/H] = -0.5, +0.3, and +0.5 (using \citet{2009ARA&amp;A..47..481A} abundances). We found a mean error of 100K, 85K, and 95K for each of the metallicities respectively. The temperatures derived using the different metallicities differed systematically from that of the temperatures derived from the [M/H]=0.0 models by -40K, +20K, and +60K respectively.

\begin{figure*}
  \centering
  \includegraphics[width=18cm]{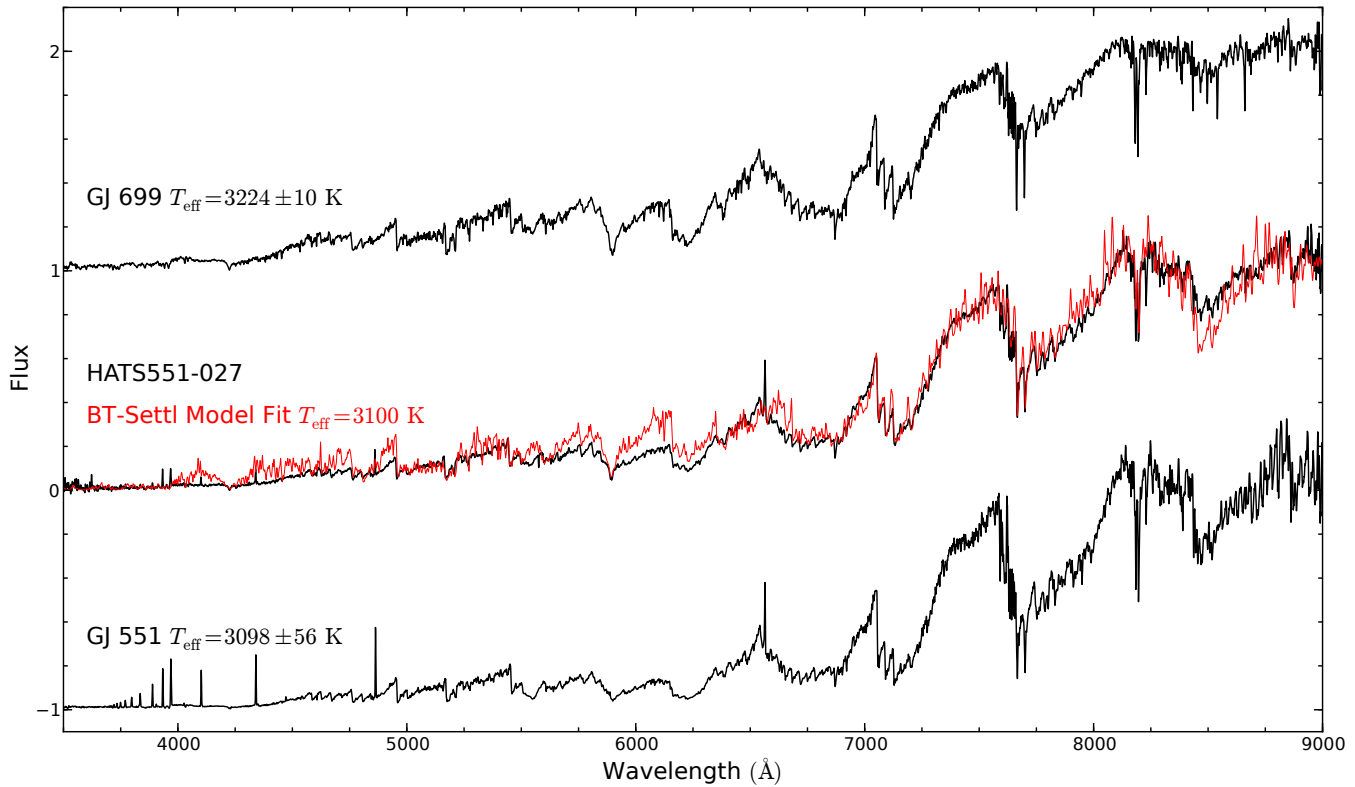}
  \caption{WiFeS $R=3000$ spectra of HATS551-027 and two standard M-dwarfs of similar a spectral type, GJ 551 and GJ 699. The observations are plotted in black, best fit BT-Settl spectral model to HATS551-027 in red.}
  \label{fig:wifes_spec}
\end{figure*}

For HATS551-027, we find a best fit binary effective temperature of $T_\text{eff}=3113\,\text{K}$. We discuss metallicity later in the section. The  WiFeS spectrum and the best fit model are plotted in Figure~\ref{fig:wifes_spec}. 

\begin{table}
  \centering
  \caption{WiFeS M-dwarf Spectral Classifications}
  \label{tab:wifes_mdwarf_char}
  \begin{tabular}{lrrc}
    \hline\hline
    Object & Literature \teff (K) & WiFeS \teff (K)$^a$ & Ref\\
    \hline
    GJ 191 & $3570\pm156$ & 3787 & 1\\
    GJ 551 & $3098\pm56$ & 3000 & 2\\
    GJ 699 & $3224\pm10$ & 3305 & 3\\
    GJ 880 & $3713\pm11$ & 3671 & 3\\
    GJ 887 & $3797\pm45$ & 3724 & 2\\
    \hline
  \end{tabular}
\begin{flushleft} 
References 1. \citet{2003A&amp;A...397L...5S} 2. \citet{2009A&amp;A...505..205D} 3. \citet{2012ApJ...757..112B}\\
$^a$ Matched against BT-Settl spectral models using \citet{2009ARA&amp;A..47..481A} abundances, with [M/H]=0.0\\
\end{flushleft}
\end{table}

For comparison, we also plot in Figure~\ref{fig:wifes_spec} the WiFeS spectra of GJ 551 \citep[M5.5V $3098\pm56\,\text{K}$,][]{2009A&amp;A...505..205D} and GJ 699 \citep[M4V $3224\pm10\,\text{K}$,][]{2012ApJ...757..112B}, two M-dwarfs with similar temperatures and spectra as HATS551-027. A visual inspection of some temperature sensitive features, such as the depth of the TiO absorption bands, the Na and CaOH lines, confirms that HATS551-027 has a spectral type between the GJ 551 and GJ 699. 

To estimate the temperatures of the two stellar components, we simultaneously fitted for the temperatures of both stellar components using a grid of simulated composite spectra. We generated a 2D grid of composite spectra from the BT-Settl models at steps of 100\,K for both simulated stellar components, with fluxes scaled using the best-fit radius ratio, and wavelength shifted in velocity according to the Keplerian orbit solution (both derived in Section~\ref{sec:glob-modell-syst}). We then calculate the \chisq of fit to the WiFeS spectrum and WiFeS CCF-derived light ratios in the R7000 and I7000 spectral bands (Section~\ref{sec:anu-2.3m}). The best fit component temperatures were $T_{\text{eff},1}=T_{\text{eff},2}=3100$, consistent with the binary effective temperature.

We can also approximate the temperature of the two stellar components by assuming they follow the Stefan-Boltzmann law. We adopt a binary effective temperature of $3114\pm85$\,K, light ratio of $0.50_{_-0.04}^{+0.10}$, and the stellar radii from Table~\ref{tab:sys_param}. The derived temperature ratio is $0.94\pm0.04$, with the temperature of the primary stellar component as $T_{\text{eff},1} = 3190\pm100$\,K and the secondary component as $T_{\text{eff},2}=2990\pm110$\,K. We adopt these temperature ratio derived temperatures for the further analysis.

\subsubsection{Metallicity}
\label{sec:metallicity}

Accurate metallicities for M-dwarfs are difficult to measure from optical spectra, since the molecular absorption features distort the continuum significantly, and mask out the common metal lines. We first use the $\zeta$TiO/CaH index to estimate the metallicity of HATS551-027. We measure the band indices for TiO5, CaH2, and CaH3, according to the regions defined in \citet{1995AJ....110.1838R}, to be 0.36, 0.39, and 0.68 respectively. We calculate a series of $\zeta$ values using relations between CaH2+CaH3 and the [TiO5]$_{Z\odot}$ index from calibrations in \citet{2007ApJ...669.1235L,2012AJ....143...67D,2013AJ....145..102L}, all yielding $\zeta=1.0$, the average abundance for disk stars. The derived metallicity estimate for HATS551-027 is $\text{[M/H]} = 0.0\pm0.2$ using the calibrations and uncertainties between $\zeta$ and [M/H] from \citet{2013AJ....145...52M}. For a consistency check, we also use the metallicity, $J-K$ and $V-K$ colour relationships from \citet{2013AJ....145...52M} to calculate a photometric colour-derived metallicity of $\text{[M/H]} = +0.1\pm0.2$, consistent with the band indices result. However, we note that infrared spectroscopy will be necessary to provide a definitive measurement for the metallicity of HATS551-027.

\subsubsection{Activity and \ha Emission}
\label{sec:ha-emission}

The low resolution WiFeS spectrum of HATS551-027 shows all the Balmer lines, and the Calcium H \& K lines, in emission. The strength of \ha emission is commonly interpreted as an indicator for stellar activity, and potentially correlated to the inflated radii of M-dwarfs \citep{2007ApJ...660..732L,2012ApJ...756...47S}. We use the PFS spectra to measure the equivalent width of the \ha emission for both stellar components of the binary, fitting Gaussian profiles to each feature. We calculate the equivalent width by assuming a polynomial fit to the continuum flux, scaled to the relative light contribution from each stellar component ($L_2/L_1=0.56_{-0.06}^{+0.06}$ Section~\ref{sec:glob-modell-syst}). The measured equivalent widths are $2.8\pm0.2\,\AA$ and $3.6\pm0.4\,\AA$ for HATS551-027A and B, respectively. To calculate the $L_{H\alpha}/L_\text{Bol}$ ratio for each stellar component, we measure a $L_\text{Bol}$ for each stellar component by integrating under the respective best fit BT-Settl model spectra. The $L_{H\alpha}/L_\text{Bol}$ ratio is calculated by comparing the flux of the Gaussian fit to the \ha lines to the integrated fluxes of the BT-Settl models. The largest error contribution comes from the $L_\text{Bol}$ estimate, stemming from the uncertainty in the atmospheric parameters. To estimate the uncertainty, we calculate a series of $L_{H\alpha}/L_\text{Bol}$ values for each component whilst adopting a grid of spectral models, with $T_\text{eff,1}$ from 3000 to 3200\,K, and $T_\text{eff,2}$ from 2900 to 3100\,K. The standard deviation scatter in the resulting $L_{H\alpha}/L_\text{Bol}$ is adopted as the uncertainty. The derived $L_{H\alpha}/L_\text{Bol}$ are $-4.15\pm0.03$ and $-4.08\pm0.05$ for the two stellar components.\\

\input{phot_params.tex}

The set of stellar photometric and spectroscopic parameters for HATS551-027 are presented in Table~\ref{tab:stellar_param}.

\section{Discussion}
\label{sec:discussion}

\begin{figure*}
  \centering
  \includegraphics[width=15cm]{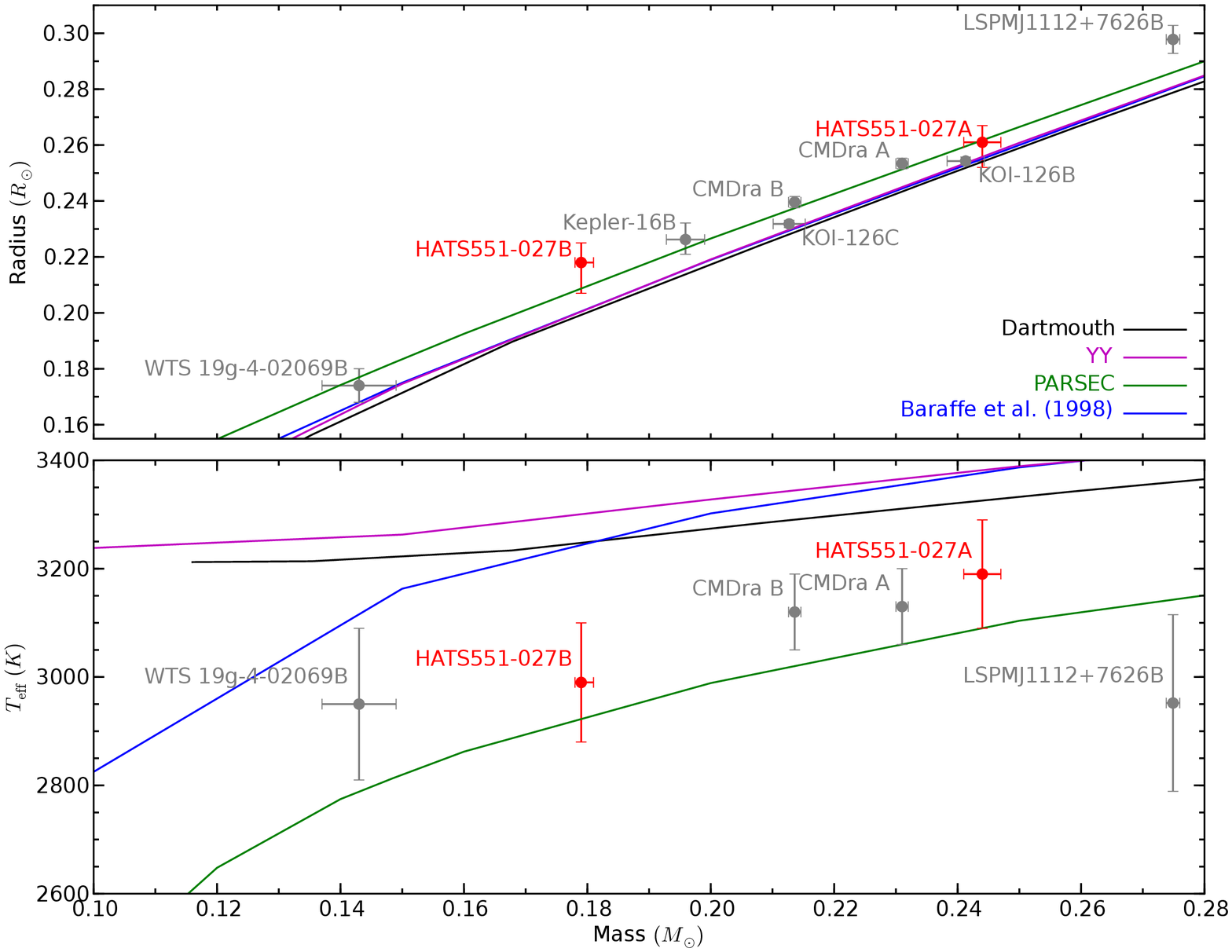}
  \caption{The measured masses, radii, and temperatures of the well characterised double-line eclipsing binaries with respect to the model isochrones. Solar metallicity, 5 Gyr models from Dartmouth \citep{2008ApJS..178...89D} are plotted in black, Yonsei-Yale \citep{2001ApJS..136..417Y,2013ApJ...776...87S} in magenta, PARSEC \citep{2012MNRAS.427..127B,2014MNRAS.444.2525C} in green, and \citep{1998A&amp;A...337..403B} in blue. Stellar properties for CM Draconis are taken from \citet{2009ApJ...691.1400M}, KOI-126 BC from \citet{2011Sci...331..562C}, the radii of Kepler-16 B from \citet{2011Sci...333.1602D} and dynamical masses from \citet{2012ApJ...751L..31B}, LSPM1112+7626 B from \citet{2011ApJ...742..123I}, and WTS 19g-4-02069 B from \citet{2013MNRAS.tmp.1072N}.}
  \label{fig:hats551-027-iso}
\end{figure*}

Following CM Draconis and KOI-126, HATS551-027 is the third well characterised double lined eclipsing binary with both components in the fully convective regime. Figure~\ref{fig:hats551-027-iso} shows the masses, radii, and temperatures of the HATS551-027 system with respect to other well characterised double-lined eclipsing binaries in the same mass regime. Figure~\ref{fig:hats551-027-iso} also compares the properties of HATS551-027 to isochrones from the Dartmouth \citep{2008ApJS..178...89D}, \citet{1998A&amp;A...337..403B}, PARSEC models \citep{2012MNRAS.427..127B,2014MNRAS.444.2525C}, and Yonsei-Yale \citep[YY,][]{2001ApJS..136..417Y,2013ApJ...776...87S} models.

The radius and estimated temperatures of these binaries are largely discrepant to the Dartmouth, YY, and \citet{1998A&amp;A...337..403B} models, but generally agree with the PARSEC model predictions to $\sim 1\sigma$. Compared to the Dartmouth 5 Gyr isochrones, the observed radii for the HATS551-027 system are larger by $2.6\pm3.5$\% and $9.1\pm4.8$\% for components A and B respectively; the temperatures are cooler by $3.8\pm3.6$\% and $7.9\pm3.7$\% respectively. The measured system properties are more consistent to the PASCAR 5 Gyr isochrones, the observed radius of component A is smaller than the models by $0.6\pm3.6$\%, and component B larger than the model by $4.2\pm4.9$\%, both consistent within errors. The temperatures for both components are also consistent to within errors, with A and B hotter than the models by $3.1\pm 5.0$\% and $2.6\pm5.6$\% respectively. The uncertainties are derived via a Monte Carlo exercise, drawing the observed sample from Gaussian distributions based on the measurement errors in mass, radius, and metallicity. For trials with sub-solar metallicities, we interpolate between the $\text{[M/H]}=-0.5$ and 0.0 isochrones to derive the model radius, for super-solar metallicites, we interpolate between the $\text{[M/H]}=0.0$ and $+0.5$ isochrones.

For tidally locked binaries, a shorter orbital period, and thereby rotation period, may lead to more powerful internal dynamos. In turn, the higher stellar activity may result in a greater discrepancy between the observed properties of these binaries and their model predictions. \citet{1989ApJ...340.1011F} established a rotation -- X-ray activity correlation for main sequence single stars. \citet{2007ApJ...660..732L} and \citet{2012ApJ...756...47S} demonstrated a correlation between X-ray activity and the radius, \teff discrepancy of low mass binaries. 

HATS551-027 has a significantly longer period (4.1 days) than both CM Draconis and KOI-126 BC ($<2$\,days). However, there appears to be no direct correlation between the model discrepancy and orbital period. In the HATS551-027 system, only HATS551-027B is significantly inflated. In the short period (1.3 days) CM Draconis binary, both stellar components are discrepant from the models in radius and temperature \citep{2002ApJ...567.1140T,2012MNRAS.421.3084M,2013ApJ...776...87S,2014ApJ...789...53F}. Compared to the Dartmouth 5 Gyr isochrones, the radii of the CM Draconis stars \citep[adopting parameters from][]{2009ApJ...691.1400M} are inflated by $5.5\pm1.0$\% and $6.1\pm1.1$\%, and temperatures cooler by $8.0\pm2.3$\% and $7.6\pm2.3$\% for components A and B respectively. However, standard isochrone models can reproduce the radii of the KOI-126 BC system very well \citep[e.g.][]{2011ApJ...740L..25F,2012MNRAS.422.2255S}, even though it is a similar close-in binary system with period of 1.8 days. Compared to the Dartmouth models, KOI-126 B is negligibly smaller by $0.2\pm1.2$\%, and C is negligibly larger by $0.5\pm1.3$\%. 

A lack of correlation remains when we also consider well characterised double-line eclipsing binaries with one low mass component. LSPM J1112+7626 \citep{2011ApJ...742..123I} has a long period of $\sim 41$\,days, with both components of the binary significantly inflated in radius by $\sim5$\%, and cooler in temperature by $\sim 10$\% compared to standard models. Kepler-16, another 41-day period unequal mass binary \citep{2011Sci...333.1602D} is also inflated in comparison to models by $\sim 7$\% if we adopt the dynamical masses measured by \citet{2012ApJ...751L..31B}. This is consistent with the findings of \citet{2012ApJ...757...42F}, which showed a lack of period correlation for detached double lined eclipsing binaries. \citet{2014MNRAS.437.2831Z} also examined a larger sample of double lined and single lined binaries in $<0.3\,M_\odot$ regime, finding no correlations with period.

We also examine \ha, as a direct activity indicator, for correlations against the radius and \teff model discrepancies. The stellar components of HATS551-027 show \ha emissions and spot modulation, signatures of significant stellar activity. Both components of HATS551-027 exhibit discrepancies in the \teff with respect to the Dartmouth models. CM Draconis system is also known to be active from it spot modulations \citep{2009ApJ...691.1400M} and X-ray emission \citep{2007ApJ...660..732L}. Although we lack a directly measured \ha emission of CM Draconis, we can estimate the \ha flux by converting the X-ray fluxes of CM Draconis reported by \citet{2007ApJ...660..732L} to $\log L_{H\alpha} / L_\text{bol}$ using the empirical relationship established for field M-dwarfs in \citet{2012ApJ...756...47S}. The approximate \ha fluxes are $\log L_{H\alpha} / L_\text{bol}=-3.8_{-1.8}^{+1.2}$ for CM Draconis A, and $-3.7_{-1.8}^{+1.2}$ for B, similar to the \ha activity we measure for HATS551-027. WTS 19g-4-02069 \citep{2013MNRAS.tmp.1072N} was also reported to exhibit strong \ha emission and \teff--model discrepancy, but we could not estimate $\log L_{H\alpha}/L_\text{bol}$ value from the reports. One exception to this trend is LSPM J1112+7626B, which has no reported \ha emission \citep{2011ApJ...742..123I}, but an estimated temperature discrepant with the Dartmouth at $\sim 6$\%.

We also measure the $\log L_{H\alpha}/L_\text{bol}$ for the single stars GJ 191, GJ 551, and GJ 699 from the WiFeS spectra that we took for spectral classifications (Section~\ref{sec:temperature}). GJ 551 exhibits a distinct \ha emission feature, but we can only provide $1\sigma$ upper limits for the \ha strength of GJ 191 and GJ 699 (Table~\ref{tab:ha_emission}).

\begin{table}
  \caption{\ha emission strength measured from WiFeS spectra}
  \label{tab:ha_emission}
  \centering
  \begin{tabular}{cc}
        \hline\hline
        Object & $\log L_{H\alpha}/L_\text{bol}$\\
        \hline
        GJ191 & $<-4.69$ \\
        GJ551 & $-4.13\pm0.18$ \\
        GJ699 & $<-4.63$ \\
        HATS551-027A & $-4.15\pm0.02$ \\
        HATS551-027B & $-4.03\pm0.05$ \\
        \hline
  \end{tabular}
\end{table}

\begin{figure}
  \centering
  \includegraphics[width=8cm]{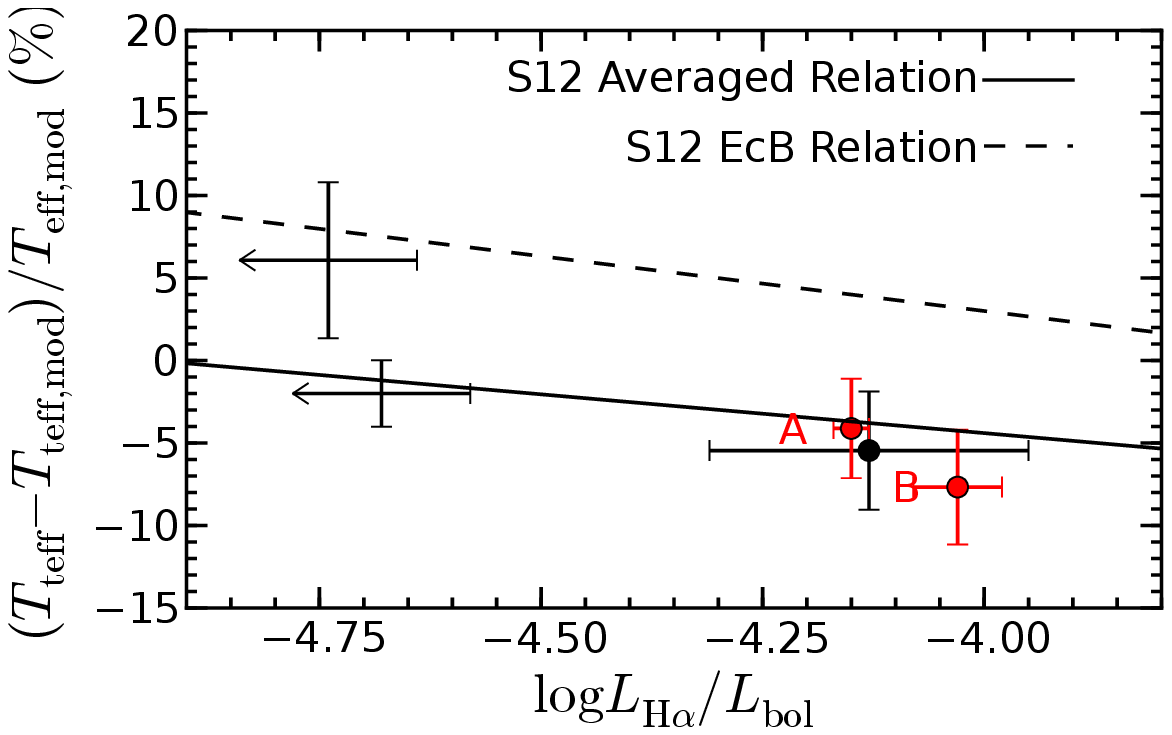}\\
  \includegraphics[width=8cm]{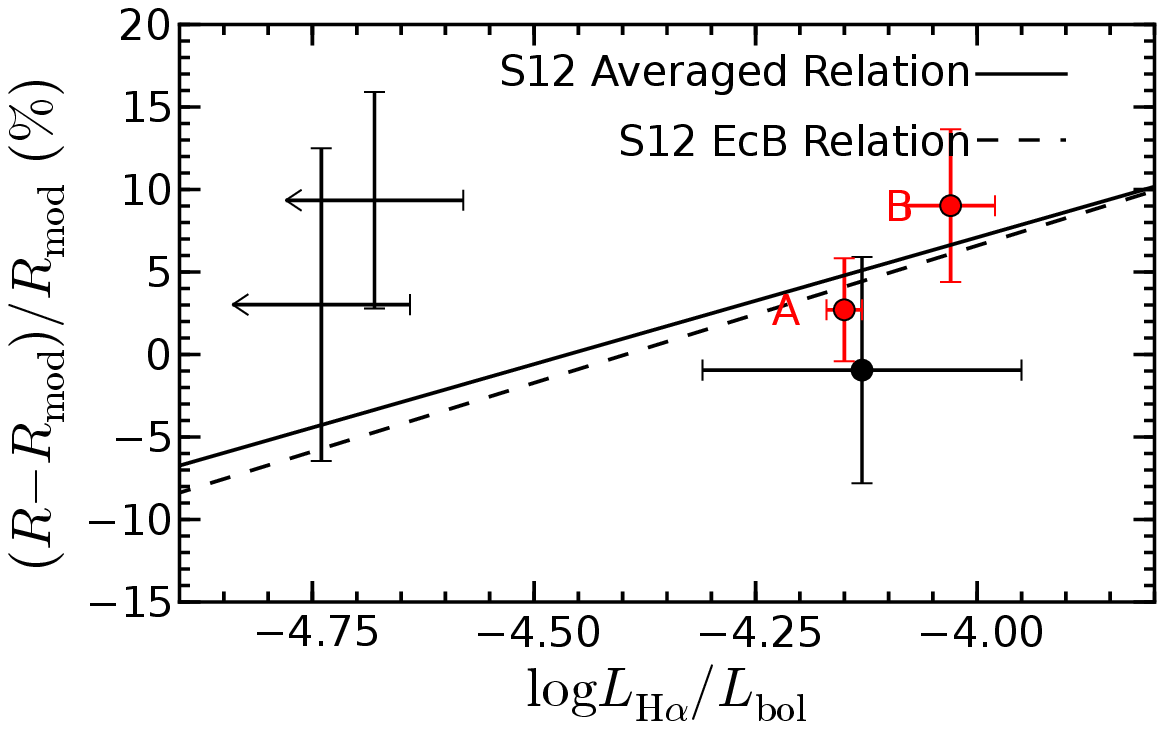}
  \caption{The \ha emission strength $(\log L_{H\alpha}/L_\text{bol})$ for the five VLMSs (Table~\ref{tab:ha_emission}) are plotted with respect to their radius and temperature deviation from the 5 Gyr solar metallicity Dartmouth isochrone. The measurements for HATS551-027 are plotted in red, with the two stellar components labelled. We also plot the empirical relationships derived by \citet{2012ApJ...756...47S} (S12 in the plot labels) using M-dwarfs $(<0.6\,M_\odot)$ with \ha emission strength derived from X-ray fluxes for comparison.}
  \label{fig:ha_correlation}
\end{figure}

Figure~\ref{fig:ha_correlation} plots the $\log L_{H\alpha}/L_\text{bol}$ values of HATS551-027 and the single stars against the temperature and radius discrepancy of the respective stars. For reference, we also plot the empirical correlations found by \citet{2012ApJ...756...47S}. Whilst the small sample size (3) prevents any meaningful statistical interpretation, we note that the \teff -- $\log L_{H\alpha}/L_\text{bol}$ relationship appears to be consistent to this analysis for the VLMSs. The two stars without apparent \ha emission, GJ 191 and GJ 699, have effective temperatures hotter than the three stars with \ha emission features, GJ 551, HATS551-027 A and B. This is potentially indicative that the M-dwarfs with cooler photospheric temperatures have hotter chromospheric temperatures. 

Given the smaller model-observation discrepancy in radius, we cannot confirm a radius--\ha activity relationship from this sample. More double-lined binaries with \ha or X-ray measurements need to be made to confirm these relationships. In addition, \ha activity cannot be the sole cause of the radius excess, given that HATS551-027A and B exhibit similar \ha activity, but only B exhibit a radius excess to the isochrones. \citet{2011ApJ...742..123I} also note a lack of detectable \ha emission features for LSPM J1112+7626, despite the inflated radii for both components of the binary.

The HATS551-027 system is also unique as the only very low mass M-M binary well positioned for southern hemisphere facilities. This should enable a host of follow-up opportunities to better characterise the properties of this benchmark system.

\section*{Acknowledgements}
\label{sec:acknowledgements}
Development of the HATSouth project was funded by NSF MRI grant
NSF/AST-0723074, operations are supported by NASA grant NNX12AH91H, and
follow-up observations receive partial support from grant
NSF/AST-1108686.
Work at the Australian National University is supported by ARC Laureate
Fellowship Grant FL0992131.
Followup observations with the MPG~2.2\,m/FEROS instrument were
performed under MPIA guaranteed time (P087.A-9014(A), P088.A-9008(A),
P089.A-9008(A)) and Chilean time (P087.C-0508(A)).
A.J. acknowledges support from project IC120009
'Millennium Institute of Astrophysics (MAS)' of the Millennium Science
Initiative, Chilean Ministry of Economy, FONDECYT project 1130857 and
BASAL CATA PFB-06.
V.S.\ acknowledges support form BASAL CATA PFB-06. R.B.\ and N.E.\
acknowledge support from CONICYT-PCHA/Doctorado Nacional and Fondecyt
project 1130857. KP acknowledges support from NASA grant: NNX13AQ62G
This work is based on observations made with ESO Telescopes at the La
Silla Observatory under programme IDs P087.A-9014(A), P088.A-9008(A),
P089.A-9008(A), P087.C-0508(A), 089.A-9006(A), and
We acknowledge the use of the AAVSO Photometric All-Sky Survey (APASS),
funded by the Robert Martin Ayers Sciences Fund, and the SIMBAD
database, operated at CDS, Strasbourg, France.
Operations at the MPG/ESO 2.2\,m Telescope are jointly performed by the
Max Planck Gesellschaft and the European Southern Observatory.  The
imaging system GROND has been built by the high-energy group of MPE in
collaboration with the LSW Tautenburg and ESO\@. We thank R\'{e}gis Lachaume for his technical assistance during the
observations at the MPG 2.2\,m Telescope.
Australian access to the Magellan Telescopes was supported through the
National Collaborative Research Infrastructure Strategy of the
Australian Federal Government.
We thank Albert Jahnke, Toni Hanke (HESS), Peter Conroy (MSO) for
their contributions to the HATSouth project.
%% EOF Acknowledgements

\bibliographystyle{mn2e}
\bibliography{mybibfile}

\FloatBarrier
%\clearpage

%\appendix
%\FloatBarrier

\label{lastpage}

\end{document}

%% file: data/lc.tex
\begin{table*}
  \centering
  \caption{HATS551-027 differential photometry}
  \label{tab:phot_table}

  \begin{tabular}{rrrll}
    \hline\hline
    BJD & Flux & $\Delta$ Flux & Instrument & Filter \\
    \hline
    &&\\
2455083.75892 & 0.9999 & 0.0236 & HATSouth$^a$ & $r'$\\
2455083.76216 & 1.0084 & 0.0192 & HATSouth & $r'$\\
2455083.76552 & 1.0514 & 0.018 & HATSouth & $r'$\\
2455083.76873 & 0.9989 & 0.0209 & HATSouth & $r'$\\
2455083.7721 & 0.9632 & 0.0221 & HATSouth & $r'$\\
2455083.77532 & 0.9487 & 0.0204 & HATSouth & $r'$\\
2455083.77868 & 1.0744 & 0.0173 & HATSouth & $r'$\\
2455083.78191 & 0.9683 & 0.02 & HATSouth & $r'$\\
2455083.78528 & 1.0118 & 0.0196 & HATSouth & $r'$\\
2455083.78849 & 1.014 & 0.0188 & HATSouth & $r'$\\
    \hline
  \end{tabular}
\begin{flushleft} 
Note. -- This table is available in a machine-readable form in the online journal.
A portion is shown here for guidance regarding its form and content.\\
$^a$For the HATSouth light curve these magnitudes have been detrended using
the EPD and TFA procedures prior to fitting a transit model to the light curve.
Primarily as a result of this detrending, but also due to blending from neighbors, the
apparent HATSouth transit depth is $\sim90$\% that of the true depth in the Sloan $r$
filter. 
\end{flushleft}
\end{table*}

%% file: data/rv.tex
\begin{table*}
  \centering
  \caption{HATS551-027 Radial Velocities}
  \label{tab:rv_table}

  \begin{tabular}{rrrrrl}
    \hline\hline
    HJD & RV$_1$ (\kms) $^a$ & $\sigma$RV$_1$ (\kms) & RV$_2$ (\kms) & $\sigma$RV$_2$ (\kms)  & Instrument\\
    \hline
    &&\\
2456179.31471 & -0.8 & 6.9 & 103.7 & 11.3 & WiFeS (\ha)\\
2456179.31471 & -25.8 & 3.3 & 70.9 & 6.7 & WiFeS (CCF)\\ %$^b$\\
2456180.28765 & 41.3 & 10.5 & 35.5 & 14.6 & WiFeS (\ha)\\
2456198.23861 & 79.5 & 12.1 & 32.0 & 19.7 & WiFeS (\ha)\\
2456198.23861 & 55.1 & 2.5 & -9.9 & 4.7 & WiFeS (CCF)\\
2456231.09138 & 44.7 & 13.5 & 32.3 & 17.4 & WiFeS (\ha)\\
2456375.98307 & 50.9 & 9.3 & 28.9 & 14.2 & WiFeS (\ha)\\
2456378.02998 & 51.3 & 10.4 & 31.8 & 16.2 & WiFeS (\ha)\\
2456378.97110 & -1.6 & 3.1 & 100.4 & 4.6 & WiFeS (\ha)\\
2456378.97110 & -18.9 & 2.1 & 74.3 & 4.3 & WiFeS (CCF)\\
2456380.99658 & 84.5 & 6.0 & -18.4 & 11.4 & WiFeS (\ha)\\
2456380.99658 & 66.2 & 2.1 & -26.8 & 6.2 & WiFeS (CCF)\\
2456876.30644 & 2.4 & 3.1 & 102.0 & 6.6 & WiFeS (\ha)\\
2456876.30644 & -18.3 & 2.8 & 75.3 & 5.1 & WiFeS (CCF)\\
2456877.30697 & 46.9 & 11.3 & 40.0 & 16.7 & WiFeS (\ha)\\
2456878.31132 & 85.1 & 7.3 & -9.8 & 12.1 & WiFeS (\ha)\\
2456878.31132 & 63.3 & 4.8 & -26.8 & 6.2 & WiFeS (CCF)\\
2456879.30918 & 48.7 & 14.2 & 38.1 & 18.7 & WiFeS (\ha)\\
2456880.30174 & 7.1 & 2.9 & 101.2 & 5.5 & WiFeS (\ha)\\
2456880.30174 & -15.8 & 1.7 & 76.6 & 4.7 & WiFeS (CCF)\\
2456881.28823 & 50.1 & 14.9 & 48.1 & 22.5 & WiFeS (\ha)\\
2456883.25386 & 56.3 & 16.4 & 19.8 & 24.8 & WiFeS (\ha)\\
2456883.25386 & 50.0 & 1.6 & -0.5 & 5.6 & WiFeS (CCF)\\
2456884.25377 & 9.5 & 4.5 & 95.6 & 6.7 & WiFeS (\ha)\\
2456884.25377 & -8.5 & 1.6 & 71.5 & 3.8 & WiFeS (CCF)\\
2456375.53799 & -5.8 & 0.2 & 65.7 & 0.6 & FEROS\\
2456376.54952 & 53.8 & 0.2 & -16.1 & 3.7 & FEROS\\
2456377.53960 & 56.6 & 0.3 & -20.1 & 4.8 & FEROS\\
2456378.52904 & -1.5 & 0.3 & 60.4 & 0.5 & FEROS\\
2456344.60054 & 3.2 & 0.3 & -93.8 & 0.9 & PFS \\
    \hline
  \end{tabular}
\begin{flushleft} 
$^a$ The zero-point of these velocities is arbitrary. An offset is added to each instrument independently in the global fitting.\\
%$^b$ The zero-point of the WiFeS CCF velocities are offset from those
%derived from the \ha line.

\end{flushleft}
\end{table*}

%% file: system_parameters.tex
\begin{table*}
  \centering
  \caption{HATS551-027 System Modelling Parameters}
  \label{tab:sys_param}

  \begin{tabular}{lrl}
    \hline\hline
    Parameter & Value & Source\\
    \hline
    % &&\\
    % RA (J2000 HH:MM:SS.SS) & 05:44:57.92 & 2MASS\\
    % DEC (J2000 DD:MM:SS.SS) & -24:56:09.6 & 2MASS\\
    % Proper Motion (mas/yr RA DEC) & 229 -25 & $^a$ \\
    % Proper Motion Error Ellipse &&\\ (mas/yr mas/yr Position Angle) & 5 5 104 & $^a$\\\
    % &&\\
    %  \multicolumn{3}{l}{\emph{Photometric parameters}}\\
    %  $B$ (mag) & $17.3\pm0.1$ & APASS\\
    %  $g'$ (mag) & $16.3$ & APASS\\
    %  $V$ (mag) & $15.54\pm0.05$ & APASS\\
    %  $r'$ (mag) & $15.0$ & APASS\\
    %  $i'$ (mag) & $13.221\pm0.05$ & APASS\\
    %  $J$ (mag) & $10.75\pm0.02$ & 2MASS\\
    %  $H$ (mag) & $10.13\pm0.02$ & 2MASS\\
    %  $K$ (mag) & $9.85\pm0.02$ & 2MASS\\
    %  &&\\
 
     \multicolumn{3}{l}{\emph{Modelled system parameters}}\\
    Period (Days) & $4.077017\,\left(_{-1}^{+1} \right)^a$ & GM$^b$\\
    $T_0$ (BJD) & $2456374.0171\,\left(_{-2}^{+1} \right)$ & GM\\
    $(R_1+R_2)/a$ & $0.0590\,\left(_{-3}^{+5} \right)$ & GM\\
    $R_2/R_1$ & $0.82 \,\left(_{-3}^{+5} \right)$ & GM\\
    $i\,(^\circ)$ & $87.99 \,\left(_{-3}^{+3} \right)$ & GM\\
    $L_2/L_1$ (6114--6574\,\AA) & $0.56 \,\left(_{-8}^{+6} \right)$ & GM+PFS\\
    $L_2/L_1$ ($R$-band) & $0.53 \,\left(_{-6}^{+6} \right)$ & WiFeS\\
    $L_2/L_1$ ($I$-band) & $0.50 \,\left(_{-1}^{+1} \right)$ & WiFeS\\
    $e\,\cos\omega$ & $0.00013 \,\left(_{-6}^{+7} \right)$ & GM\\
    $e\,\sin\omega$ & $-0.002 \,\left(_{-2}^{+1} \right)$ & GM\\
    $K_1\,(\text{km\,s}^{-1})$ & $42.4 \,\left(_{-1}^{+2} \right)$ & GM\\
    $K_2\,(\text{km\,s}^{-1})$ & $57.6 \,\left(_{-3}^{+4} \right)$ & GM\\
    $R$-band star A $u_{1}$ & 0.3762 & Fixed$^c$\\
    $R$-band star A $u_{2}$ & 0.4574 & Fixed$^c$\\
    $R$-band star B $u_{1}$ & 0.5392 & Fixed$^c$\\
    $R$-band star B $u_{2}$ & 0.3485 & Fixed$^c$\\
    $i$-band star A $u_{1}$ & 0.1539 & Fixed$^c$\\
    $i$-band star A $u_{2}$ & 0.6330 & Fixed$^c$\\
    $i$-band star B $u_{1}$ & 0.3124 & Fixed$^c$\\
    $i$-band star B $u_{2}$ & 0.5509 & Fixed$^c$\\
    &&\\
    \multicolumn{3}{l}{\emph{Gaussian process modelling parameters}}\\
    $A$ (FTS 20121212) & $0.0005  \,\left(_{-2}^{+2} \right)$& GM\\
    $\tau$ (FTS 20121212) (days) & $0.0013  \,\left(_{-1}^{+1} \right)$ &GM\\
    $A$ (Swope 20130226) & $0.0022  \,\left(_{-5}^{+6} \right)$ &GM\\
    $\tau$ (Swope 20130226) (days) & $0.0037  \,\left(_{-2}^{+4} \right)$ &GM\\
    $A$ (FTS 20130320) & $0.0004  \,\left(_{-4}^{+3} \right)$ &GM\\
    $\tau$ (FTS 20130320) (days) & $0.0019  \,\left(_{-2}^{+1} \right)$ &GM\\
    &&\\
     \multicolumn{3}{l}{\emph{Derived parameters}}\\
    $M_1\,(M_\odot)$ & $0.244  \,\left(_{-3}^{+3} \right)$ & GM\\
    $R_1\,(R_\odot)$ & $0.261 \,\left(_{-9}^{+6} \right)$ & GM\\
    $M_2\,(M_\odot)$ & $0.179 \,\left(_{-1}^{+2} \right)$ & GM\\
    $R_2\,(R_\odot)$ & $0.218  \,\left(_{-11}^{+7} \right)$ & GM\\
    %$L_2/L_1$ ($r$-band) & $0.50 \,\left(_{-4}^{+10} \right)$ & GM+PFS+BT-Settl\\
    %$L_2/L_1$ ($i$-band) & $0.50 \,\left(_{-4}^{+10} \right)$ & GM+PFS+BT-Settl\\

    \hline
  \end{tabular}
\begin{flushleft} 
%$^a$ \citet{2003ApJ...582.1011S}\\
$^a$ Uncertainties in parenthesis for the last significant figure\\
$^b$ GM - Global modelling of light curves and radial velocities\\
$^c$ Fixed to values interpolated from \citet{2000A&amp;A...363.1081C}
using the PHOENIX models.
\end{flushleft}
\end{table*}

%% file: phot_params.tex
\begin{table*}
  \centering
  \caption{HATS551-027 Stellar Parameters}
  \label{tab:stellar_param}

  \begin{tabular}{lrl}
    \hline\hline
    Parameter & Value & Source\\
    \hline
    &&\\
    RA (J2000 HH:MM:SS.SS) & 05:44:57.92 & 2MASS\\
    DEC (J2000 DD:MM:SS.SS) & -24:56:09.6 & 2MASS\\
    Proper Motion (mas/yr RA DEC) & 229 -25 & \citet{2003ApJ...582.1011S} \\
    Proper Motion Error Ellipse &&\\ (mas/yr mas/yr Position Angle) & 5 5 104 &\citet{2003ApJ...582.1011S}\\\
    &&\\
     \multicolumn{3}{l}{\emph{Photometric parameters}}\\
     $B$ (mag) & $17.3\pm0.1$ & APASS\\
     $g'$ (mag) & $16.3$ & APASS\\
     $V$ (mag) & $15.54\pm0.05$ & APASS\\
     $r'$ (mag) & $15.0$ & APASS\\
     $i'$ (mag) & $13.221\pm0.05$ & APASS\\
     $J$ (mag) & $10.75\pm0.02$ & 2MASS\\
     $H$ (mag) & $10.13\pm0.02$ & 2MASS\\
     $K$ (mag) & $9.85\pm0.02$ & 2MASS\\
    &&\\
    \multicolumn{3}{l}{\emph{Spectroscopic parameters}}\\
    %$T_\text{eff,binary}$ (K) & $3114\pm85$ & WiFeS\\
    %$T_\text{eff,1} = T_\text{eff,2}$ (K) & $3114\pm85$ & WiFeS\\
    $T_\text{eff,1}$ (K) & $3190\pm100$ & WiFeS\\
    $T_\text{eff,2}$ (K) & $2990\pm110$ & WiFeS\\
    $\text{[M/H]}$ & $0.0\pm0.2$ & WiFeS\\
    $\log L_{H\alpha,1}/L_\text{bol,1}$ & $-4.15\pm0.03$ & PFS\\
    $\log L_{H\alpha,2}/L_\text{bol,2}$ & $-4.08\pm0.05$ & PFS\\
    \hline
  \end{tabular}
%\begin{flushleft} 
%$^a$ \citet{2003ApJ...582.1011S}\\
%\end{flushleft}
\end{table*}